\documentclass[pre,twocolumn,notitlepage,longbibliography,amsmath,amssymb,floats,superscriptaddress,nofootinbib,10pt]{revtex4-1}

\pdfoutput=1

\AtBeginDocument{%
    \newwrite\bibnotes
    \def\bibnotesext{Notes.bib}
    \immediate\openout\bibnotes=\jobname\bibnotesext
    \immediate\write\bibnotes{@CONTROL{REVTEX41Control}}
    \immediate\write\bibnotes{@CONTROL{%
    apsrev41Control,author="08",editor="1",pages="1",title="0",year="1"}}
     \if@filesw
     \immediate\write\@auxout{\string\citation{apsrev41Control}}%
    \fi
}%

\usepackage{graphicx}
\usepackage{dcolumn}
\usepackage{bm}
\usepackage{revsymb}
\usepackage[usenames]{color}
\usepackage{color}
\usepackage{physics}
\usepackage{amsmath}
\usepackage{cancel}
\usepackage[caption=false]{subfig}

\usepackage{amsfonts}
\usepackage{amssymb,bbm}
\usepackage{soul}

\newcommand{\vct}[1]{\mathbf{#1}}

\usepackage{xcolor}

\newcommand{\bea}{\begin{eqnarray}}

\definecolor{nblue}{RGB}{28,130,185}

\definecolor{cgreen}{RGB}{76,153,0}

\definecolor{myorange}{RGB}{245,156,74}

\usepackage{hyperref}
\hypersetup{
  colorlinks=true,
  citecolor=magenta,
  urlcolor=-myorange
}

\newcommand{\quotes}[1]{``#1''}

\newcommand{\ull}[1]{\underline{\underline{
#1}}}

\usepackage{xcolor}
\definecolor{ogreen} {RGB}{71,191,145}
\definecolor{bblue} {RGB}{137,207,240}

\definecolor{edit} {RGB}{123,150,145}
\definecolor{purple} {RGB}{148,0,211}

\definecolor{oblue} {RGB}{50,50,250}

\begin{document}

\title{Molecular modelling of odd viscoelastic fluids}
\begin{abstract}
We consider an active, stochastic microscopic model of particles suspended in a fluid and show that the coarse-grained description of this model renders odd viscoelasticity. The particles are odd dumbbells, each featuring a robotic device as the bead, which exhibits a particular torque response. We show that this model can be macroscopically treated as a viscoelastic fluid, analytically calculate the coefficients of the corresponding viscoelastic model, and corroborate the results using molecular dynamics simulations. This work provides a unified analytical framework for several experimental and numerical setups designed to elucidate odd effects in fluids.
\end{abstract}

\author{Paweł Matus}
\email{matus@pks.mpg.de}
\affiliation{Max Planck Institute for the Physics of Complex Systems and W\"urzburg-Dresden Cluster of Excellence ct.qmat, 01187 Dresden, Germany}

\author{Ruben Lier}
\email{r.lier@uva.nl}
\affiliation{Institute for Theoretical Physics, University of Amsterdam, 1090 GL Amsterdam, The Netherlands}
\affiliation{Dutch Institute for Emergent Phenomena (DIEP), University of Amsterdam, 1090 GL Amsterdam, The Netherlands}

\author{Piotr Sur\'{o}wka}
\email{piotr.surowka@pwr.edu.pl}
\affiliation{Institute of Theoretical Physics, Wroc\l{}aw  University  of  Science  and  Technology,  50-370  Wroc\l{}aw,  Poland}

\maketitle

\section{Introduction}

Active matter is a class of matter that is composed of individual agents that can consume energy from their environment to perform work \cite{PhysRevX.12.010501,RevModPhys.85.1143}. Unlike passive systems that primarily respond to external stimuli, components of active matter are capable of self-propulsion and can exhibit coordinated behavior. Examples include biological systems like flocks of birds, schools of fish, cells within living organisms, as well as synthetic systems like self-propelled colloidal particles or ensembles of robots. When applied to metamaterials, activity introduces a host of new capabilities and features that make these materials even more intriguing for future technologies \cite{Brandenbourger2019,PhysRevLett.121.178001}. 

Odd active materials represent a fascinating intersection of active matter and systems with broken chiral symmetry. These materials manifest as odd fluids and odd crystals, characterized by their odd viscous or odd elastic response, respectively \cite{Fruchart_2023}. Odd elastic materials exist in the form of chiral starfish embryo crystals \cite{Tan2022,Bililign_2021} or metamaterials that are networks of non-reciprocal springs \cite{Scheibner2020,Chen2021}. Odd fluid behavior has been observed experimentally and numerically for suspensions of externally rotated spinners \cite{Soni2019,Han2021}, but also for passive systems \cite{KORVING19665,berdyugin2019measuring} as, unlike odd elasticity, odd viscosity does not require activity to manifest.

Simultaneously, various common materials such as rubbers, wood, ligaments and tendons, or metals near their melting point can display both solid-like and fluid-like properties depending on the timescale of the experiment, a phenomenon known as viscoelasticity \cite{Lakes_2009}. This observation suggests that the behavior of parity-breaking materials can be more generically described by a novel hydrodynamic theory dubbed “odd viscoelasticity" \cite{Banerjee_2021,lier2021passive,floyd2022signatures,surowka_2022_cosserat,duclut2023probe,PhysRevE.105.064603}. Odd viscoelastic theories feature unconventional transport coefficients, signatures of which may include unusual pattern formation \cite{floyd2022signatures} and probe particles displaying oscillatory velocity at long times \cite{duclut2023probe}. Interestingly, unlike odd elastic solids \cite{Scheibner2020}, odd viscoelastic fluids can theoretically be found in active and passive systems \cite{lier2021passive}.

When formulating a theory of viscoelasticity, one typically starts with a macroscopic description in terms of stress, strain and its derivatives. However, understanding the microscopic origin of viscoelasticity is critical both at the theoretical as well as the practical level. To this end, various toy systems can be employed as a first step towards the development of a realistic model. Dumbbell models can be used to model the behavior of complex macromolecular systems like polymers in a simple way \cite{renardy,HINCH2021104668,Peterlin+1966+563+586,Bird1987-gg,larson1999structure,PhysRevLett.131.194002}. Upon coarse-graining these models, the continuum theory one arrives at is that of a viscoelastic fluid. In this approach, a polymer molecule is approximated as a dumbbell consisting of two large beads connected by a spring. The beads represent clumps of atoms or monomers, while the spring symbolizes the chain of monomers connecting them.

In this work, we consider a system of odd active agents suspended in a fluid. The odd active agents that are being considered are an active generalization of molecular dumbbell models (see also \cite{mandapuhydroequations,epsteinnemd,epsteindiff}). By coarse-graining the microscopic equations and performing numerical simulations, we show that such a system gives rise to odd viscoelasticity and we provide analytical expressions for the coefficients in the resulting model. Interestingly, similar systems have been considered both in experiment \cite{Soni2019} and numerical studies \cite{van_Zuiden_2016,epsteinnemd, epsteindiff,Han2021,PhysRevLett.130.158201,Lou2022,poggioli2023emergent}, with Refs. \cite{epsteinnemd, Han2021} observing unusual time evolution of the stress response. At the same time, to the best of our knowledge there has been no attempt at understanding the odd viscoelastic behavior by coarse-graining a microscopic model. Deriving macroscopic properties of such a model can serve as a guide when designing metamaterials with desired features and a point of reference for understanding experimental or numerical results. This is precisely the purpose of this work.

The structure of this work is as follows. In Sec. \ref{sec:springpart} we describe experimental systems of suspended dimers that display odd springs constants and formulate their corresponding microscopic equations of motion with stochastic forces. In Sec. \ref{sec:twofluid} and Sec. \ref{sec:jeffreys} we coarse-grain this model to find that at a macroscopic level it is described by a theory of odd viscoelasticity with two distinct elastic degrees of freedom with different relaxation times. Then in Sec. \ref{sec:numerical} we compute the evolution of the elastic stress as a function of time and compare the result to a numerical simulation. 

\section{Odd dumbbells}
\label{sec:springpart}

\begin{figure}
    \includegraphics[width=0.80\linewidth]{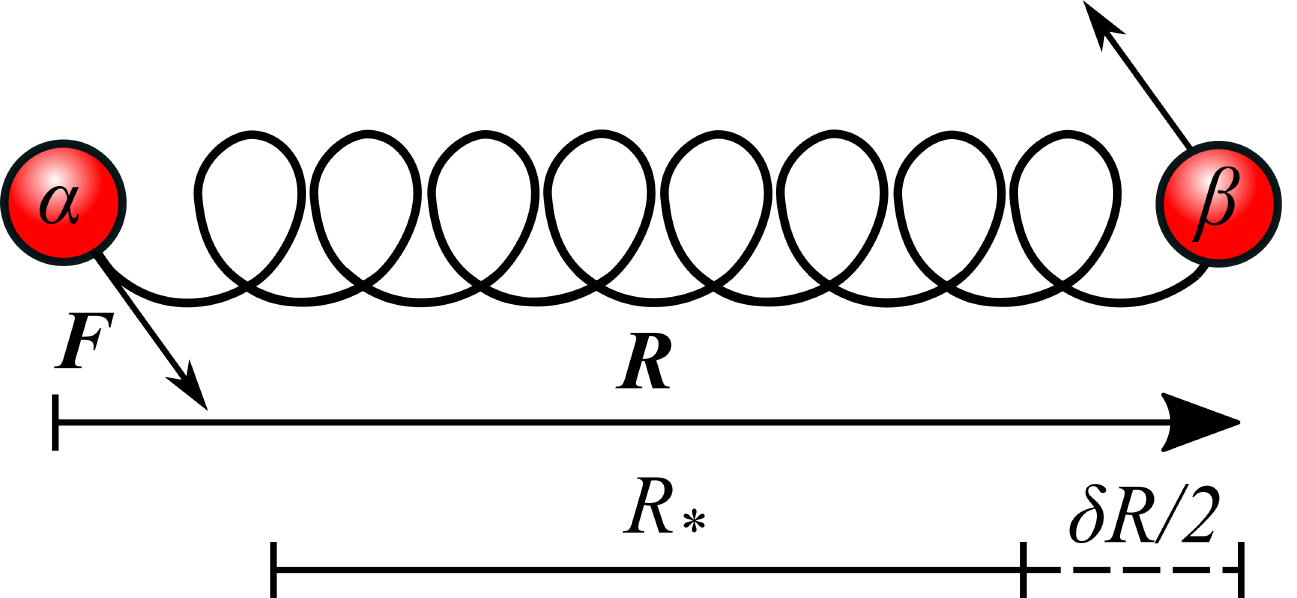}
    \caption{Dumbbell in a decompressed state compared to the rest state. Arrows represent the dumbbell forces given by Eq.~\eqref{eq:force_approx}.}
    \label{fig:my_label8u987}
\end{figure}
\begin{figure}
    \includegraphics[width=1\linewidth]{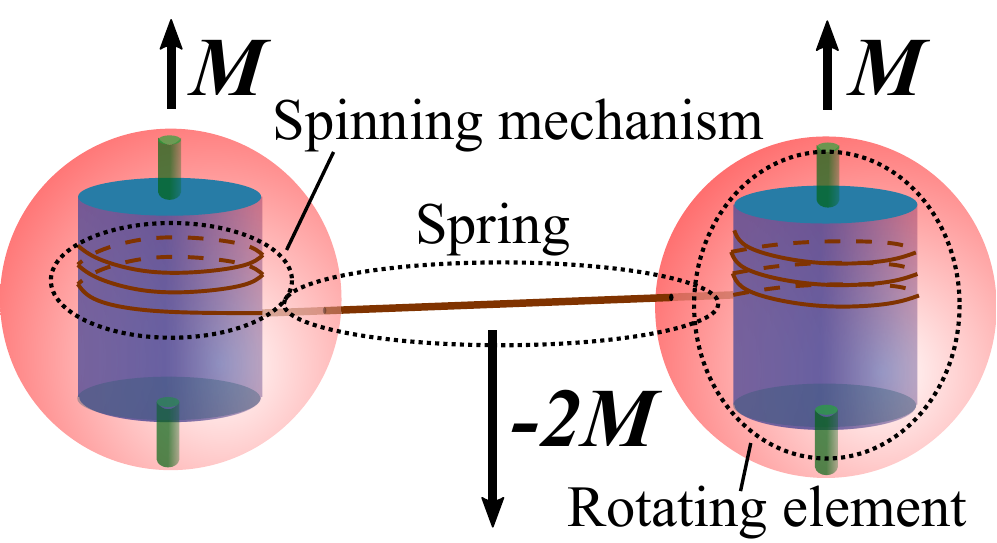}
    \caption{Schematic model of an odd dumbbell. Cylinders inside the beads represent rotating elements, such as flywheels, that act as internal storage of angular momentum. The beads are connected by an elastic element that functions as a spring. The coils wound around the cylinder represent a spinning mechanism that applies an equal torque $M$ to each of the internal rotating elements when the spring is deformed. Due to the conservation of the total angular momentum, torque equal to $-2M$ (represented by the middle arrow) has to be simultaneously experienced by the rest of the dumbbell. Thus, assuming $M\propto \delta R$, the mechanism effectively applies transverse forces $F_\mathrm{odd}\propto \delta R$ to the beads.}
            \label{fig:torque_model}
\end{figure}

We will now describe a microscopic model of active dumbbells which display both a longitudinal and transverse force when strained away from their rest state. Introducing the intra-dumbbell distance vector $\vct R$, we denote the dumbbell's length at rest by $|\vct R|_\mathrm{rest} = R^\star$ (see Fig.~\ref{fig:my_label8u987}). Let us parametrize $\vct R$ by a scalar $\delta R \equiv |\vct R|-R^*$ expressing the deviation from the rest state, and the unit vector $\hat{\vct r} \equiv \vct R/|\vct R|$. For a small deviation $\delta R$, such that $\delta R/R_* \ll 1$, the force that the odd spring exerts on a bead can be expressed as \cite{Scheibner2020}
\begin{equation}
    F_i(\vct R) \approx \delta R \left(\kappa_e \delta_{ij}+\kappa_o \varepsilon_{ij}\right) \hat{r}_j
    \label{eq:force_approx}
\end{equation}
where $i,j$ are spatial indices, $\varepsilon_{ij}$ is the Levi-Civita tensor and $\kappa_e$ and $\kappa_o$ represent the even and odd spring constants respectively. The broad features of the behaviour of the system will not depend on the exact formula for $F_i(\vct R)$ as long as Eq. (\ref{eq:force_approx}) is valid. In order to proceed with analytical treatment, we take the elastic force exerted by the spring to be given by:
\begin{equation}
     F_{i,\mathrm{elastic}} (\vct R)  = \delta R\,\kappa_e \hat{r}_i \,.  
     \label{spring_forcfheiuhe}
\end{equation}  
Because the elastic energy of the dumbbell is equal to $\kappa_e (\delta R)^2/2$, the equipartition theorem ensures that in equilibrium the average value of $\delta R^2$ is \footnote{The temperature $T$ is understood as the temperature of the bath in which the dumbbells are suspended. We assume that $T$ is not affected by the activity of the suspended dumbbells. Eq.~(\ref{eq:R2_average}) can be derived explicitly from the Fokker-Planck equation presented later on in Eq.~(\ref{fp_eq}).}
\begin{equation}
    \langle \delta R^2 \rangle = \frac{kT}{\kappa_e} \,.
    \label{eq:R2_average}
\end{equation}
Thus, the approximation (\ref{eq:force_approx}) will in general be valid when $kT/(\kappa_e R_*^2) \ll 1$, which is what we assume from now on.

One way to introduce a transverse force consistent with Eq. (\ref{eq:force_approx}) is by coupling the beads to an internal reservoir of angular momentum which interacts with the rest of the dumbbell when the bond between the beads is stretched or compressed, as visualized schematically in Fig.~\ref{fig:torque_model}. Let us first consider the possibility that the torque $M$ is directly proportional to the deformation of the spring $\delta R$. Such a mechanism is referred to as an ``odd spring" in the literature \cite{Scheibner2020}; a possible realisation of an odd spring for the dumbbells is discussed in more detail in Appendix \ref{app:oddspring}. We can then parametrize $M = \varepsilon_{ij}R_i F_j = \kappa_o R_* \delta R$ for some coefficient $\kappa_o$, from which follows that the odd part of the force is 
\begin{equation}
    F_{i,\mathrm{odd}} = \kappa_o \frac{R_*}{R}\delta R\, \varepsilon_{ij}\hat{r}_j\approx \delta R\,\kappa_o \varepsilon_{ij}\hat{r}_j   \,,
    \label{eq:f_odd_full}
\end{equation}
where we expanded in $\delta R/R_*$ and $R \equiv  |\vct R|$.

Remaining within the model of Fig.~\ref{fig:torque_model}, we can also consider the possibility that the cylinders are attached to mainsprings that are kept wound, which leads to a constant torque acting on the dumbbells independently of $\delta R$. We can include this in our description of the system by introducing a constant torque parametrized as $M=\zeta \Omega R_*^2/2$ which produces a driving force 
\begin{equation}
    F_{i,\mathrm{rot}} = \frac{\zeta\Omega}{2} ~\frac{R_*^2}{R}\varepsilon_{ij}\hat{r}_j \approx \frac{\zeta \Omega \left(R_*-\delta R\right)}{2}\,\varepsilon_{ij}\hat{r}_j \,.
    \label{force_rot}
\end{equation}
The frequency $\Omega$ corresponds to the frequency of counterclockwise dumbbell rotation in overdamped dynamics with $\zeta$ representing the particle drag. 
\begin{figure}
    \includegraphics[width=0.8\linewidth]{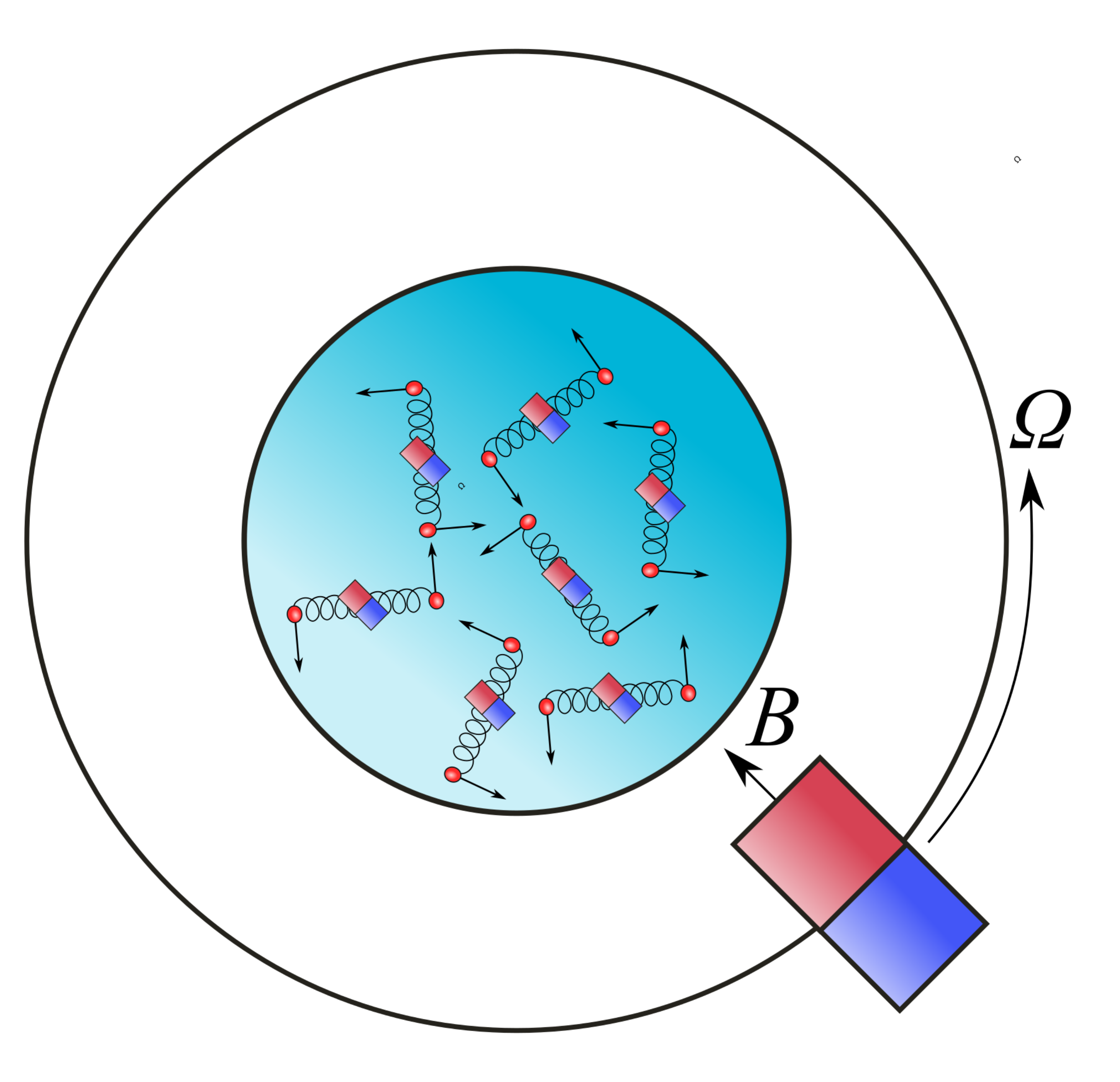}
    \caption{A rotating background magnetic field allows for springs with a magnetic moment to display an odd spring force.}
            \label{fig:my_label871}
\end{figure}

At this point we note that there can exist other mechanisms that can induce a transverse force of a form similar to that in Eq. (\ref{force_rot}). For example, the $\Omega$ term can possibly be used to describe the rotation of a suspension of particles possessing magnetic moments and subjected to a rotating magnetic field, as in Fig.~\ref{fig:my_label871}. In the latter case, the torque acting on the dumbbells was found in Ref.~\cite{Soni2019} to lock the angular velocity of the particles to the frequency $\Omega$ of the rotating magnetic field such that their rotation is independent of the fluid medium in which the particles lie. When this happens, objectivity (independence of stress from vorticity) is violated and a non-vanishing rotational viscosity $\eta_R$ appears which relaxes the fluid vorticity $\omega$ to $2 \Omega$ through adding a term proportional to $-\eta_R\left(2\Omega-\omega\right)$ to the stress tensor. As shown in Appendix~\ref{app:modelderivation}, we find $\eta_R$ to be given by the formula
\begin{equation} \label{eq:rotational}
        \eta_R = \frac{ n^\mathrm{d} R_*^2\zeta}{8}\,, 
\end{equation}
where $n^\mathrm{d}$ is the dumbbell density. We analyze the effect of the angular motion locking in more detail in Appendix \ref{app:modelderivation}, but in the main text we focus on the case without the locking.

In addition to the intra-dumbbell forces, we also consider interactions with a background fluid which damp the motion of the beads with a rate $\zeta$ and introduce a stochastic force $\vct S^{(\lambda)}$ acting on the beads. In the overdamped regime, this gives the following equations of motion:
\begin{subequations} \label{eq:stochasticstart}
\begin{align}
  \zeta \left(     \dot{\vct r}^{(\alpha)} - \vct u (\vct r ^{(\alpha)}) \right)  &=  \vct F (\vct R) + \vct S^{(\alpha )}\,, \\ 
         \zeta \left(     \dot{\vct r}^{(\beta )} - \vct u (\vct r ^{(\beta)})\right)  &= -  \vct F (\vct R) + \vct S^{(\beta)}\,, 
\end{align}
\end{subequations}
where $\vct r^{(\lambda)}$ are vectors of the coordinates of the dumbbell components, $\vct R = \vct r^{(\beta)}-\vct r^{(\alpha)}$, $\vct u$ is the velocity of the background fluid, and $\zeta$ is the drag coefficient. The stochastic forces are modelled as a white noise, i.e.,
\begin{equation}
    \langle S^{(\alpha)}_i S^{(\alpha)}_j \rangle = 2 D \zeta^2\delta_{ij}\delta(t_2-t_1)
\end{equation}
and similarly for $\vct S^{(\beta)}$, while $\langle S^{(\alpha)}_i S^{(\beta)}_j \rangle = 0$. Note that in this model the beads interact with one another only indirectly via the springs and the background fluid, while the effects of bead collisions are neglected. We also neglect short-range hydrodynamic interactions, meaning that the model is valid in the long-dumbbell limit $R_*\gg a$, where $a$ is the radius of a bead.

\section{Continuity equations}
\label{sec:twofluid}
We now turn to derive odd viscoelastic behaviour from the model introduced in Sec. \ref{sec:springpart}. For this, in addition to the relative vector $\vct R \equiv \vct r^{(\beta)}-\vct r^{(\alpha)}$ we define the center of mass $\vct x \equiv (\vct r^{(\alpha)}+\vct r^{(\beta)})/2$, so that we can turn Eq.~\eqref{eq:stochasticstart} into 
\begin{subequations}
    \begin{align}
    \dot{R}_i &= R_j \partial_j u_i - \frac{2}{\zeta} F_i (\vct R) +\frac{1}{\zeta} f^1_i\,, \\
    \dot{x}_i &= u_i + \frac{1}{\zeta}f^2_i\,, 
\end{align} \label{eq:eom}
\end{subequations}
where we defined $\vct f^1 = \vct S^{(\beta)} - \vct S^{(\alpha)}$ and $\vct f^2 = (\vct S^{(\alpha)} + \vct S^{(\beta)})/2$. Now we introduce the function $\psi(\vct x, \vct R, t)$ which represents the probability of finding a dumbbell at position $\vct x$ at time $t$ with the intra-dumbbell vector equal to $\vct R$. The Fokker-Planck equation for the evolution of $\psi(\vct x, \vct R, t)$ can be derived from Eq. (\ref{eq:eom}), and it takes the form~\cite{renardy,Bird1987-gg,larson1999structure}
  \begin{multline}
  \frac{\partial}{\partial t}\psi+ \frac{\partial}{\partial  x_i }\left( u_i  \,\psi \right) = -\partial_j u_i \frac{\partial}{\partial R_i}\left(R_j \psi\right) + \\
    + \frac{2}{\zeta}  \frac{\partial}{\partial R_i} \left( F_i (\vct R)\psi\right) + \frac{2 kT}{\zeta}\Delta_R \psi + \frac{k T}{2\zeta} \Delta_x \psi\,, 
\label{fp_eq}
\end{multline} 
where we take into account the Einstein relation $D=k T/\zeta$ with $T$ representing the temperature of the fluid bath. In Eq.~\eqref{fp_eq}, $\Delta_R$ and $\Delta_x$ are Laplace operators with respect to $R_i$ and $x_i$ respectively.

Before proceeding further, let us introduce two isotropic rank-two tensors:
\begin{subequations}
\begin{align}
    \gamma_{ijkl} & = \frac12 \delta_{ik}\delta_{jl}+\frac12 \delta_{il}\delta_{jk}-\frac12 \delta_{ij}\delta_{kl}\,,\\
    \gamma^\mathrm{o}_{ijkl} & = \frac14\left(\varepsilon_{ik}\delta_{jl}+\varepsilon_{il}\delta_{jk}+\varepsilon_{jk}\delta_{il}+\varepsilon_{jl}\delta_{ik}\right)\,.
\end{align}
\end{subequations}
So defined $\gamma_{ijkl}$ and $\gamma^\mathrm{o}_{ijkl}$ are both traceless and symmetric under the exchange of $i \leftrightarrow j$ and $k \leftrightarrow l$, but while $\gamma_{ijkl}$ is parity-even:
$\gamma_{ijkl} = \gamma_{klij}$, $\gamma^\mathrm{o}_{ijkl}$ is parity-odd: $\gamma^\mathrm{o}_{ijkl} = -\gamma^\mathrm{o}_{klij}$. We then denote for any tensor $C_{ij}$
\begin{equation}
\begin{split}
    C_{\langle i j \rangle} & = \gamma_{ijkl} C_{kl}\,,  \quad C^\mathrm{o}_{\langle i j \rangle}  = \gamma^\mathrm{o}_{ijkl} C_{kl}\,,\\ 
    C & = \Tr C_{ij}\,, \quad\quad C^A = \varepsilon_{ij}C_{ij}\,.
\end{split}
\end{equation}
As explained in App.~\ref{app:stressderivation}, we will assume the dumbbells in the fluid to be dilute, so that all the viscous and inertial contributions from the dumbbells are small compared to the fluid contributions and are thus neglected.
With these preparations, the mass and momentum conservation of the fluid can be described in terms of the following continuity equations
\begin{subequations}
\begin{align} 
    \partial_t \rho^\mathrm{f}  + \partial_i ( \rho^\mathrm{f} u_i ) &= 0\,, \\
    \rho^\mathrm{f} (  \partial_t u_i +   u_j \partial_j  u_i ) +  \partial_j T_{ij} &= 0\,, \label{eq:momentum_conserv}
    \end{align} \label{eq:continutities}
\end{subequations}
where $\rho^\mathrm{f}$ is the fluid density and the stress tensor $T_{ij}$ can be decomposed as
\begin{equation}
     T_{ij}   =  T^{(\mathrm{f})}_{ij} +T^{(\mathrm{s})}_{ij}+p^{\mathrm{d}}\delta_{ij}    \,. 
     \label{eq:T_decompose}
\end{equation}
$T^{(\mathrm{f})}_{ij} $ is the stress tensor of the fluid, which takes form
\begin{equation}
    T^\mathrm{(\mathrm{f})}_{ij} = p^\mathrm{f} \delta_{ij} - \eta_{ijkl} D_{kl}\,,
\end{equation}
where $D_{ij} = \partial_{i} u_{j}$, $p^\mathrm{f}$ is the fluid pressure and
\begin{equation}
    \eta_{ijkl} = \eta_s \gamma_{ijkl}+\frac12 \eta_b\delta_{ij}\delta_{kl}\,.
\end{equation}
Furthermore, $p^{\mathrm{d}}=2n^{\mathrm{d}}kT$ is the dumbbell pressure and $T^\mathrm{(\mathrm{s})}_{ij}$ is the component of the stress tensor related to the spring force and given by the Irving-Kirkwood virial stress formula \cite{irvingkirkwood,Bird1987-gg,renardy}
\begin{equation}   
    T^\mathrm{(\mathrm{s})}_{ij}  = -\int d^2 R~ R_j F_i(\vct R) \psi\,.
    \label{eq:Ts}
\end{equation}
The detailed derivation of Eqs.~\eqref{eq:continutities},~\eqref{eq:T_decompose},~\eqref{eq:Ts} can be found in App.~\ref{app:stressderivation}. There we show how by calculating the zeroth and first moment of the Fokker-Planck equation with respect to the dumbbell momentum one can find the continuity equations for the dumbbell density and momentum, respectively. These equations are then added to the corresponding equations for the background fluid, producing the abovementioned formulas.

\section{Odd Jeffreys model}
\label{sec:jeffreys}
The stress tensor $T_{ij}$ given in Eq.~\eqref{eq:T_decompose} can be decomposed into a constant part and a non-equilibrium part $\tilde{T}_{ij}$ as follows:
\begin{equation}
    T_{ij} = (p^\mathrm{f}+n^\mathrm{d}kT)\delta_{ij} - \frac{\kappa_o}{2\kappa_e}n^\mathrm{d}kT\varepsilon_{ij} -\frac{n^\mathrm{d} R_*^2 \zeta \Omega   }{4}\varepsilon_{ij} + \tilde{T}_{ij}\,.
    \label{eq:Ts_long_main}
\end{equation} 
The constant antisymmetric stress is proportional to the strength of the driving and expresses a steady injection of angular momentum into the system. Note that by tuning the strength of the driving to $\zeta\Omega/2 = -kT\kappa_o/(\kappa_e R_*^2)$ the constant torque can be made to vanish.

\begin{figure}
\includegraphics[width=0.8\linewidth]{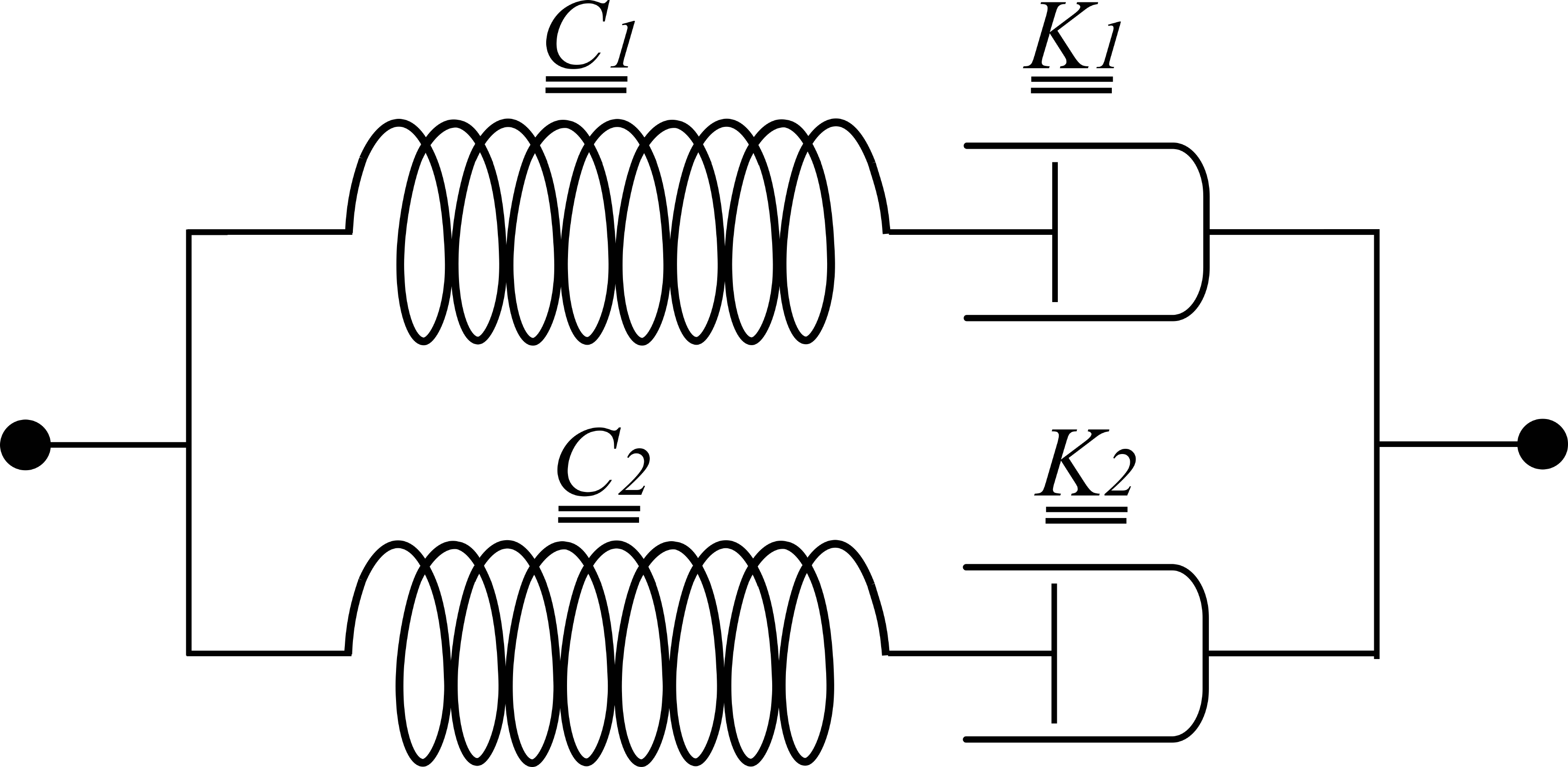}
    \caption{Viscoelastic circuit corresponding to the generalized Maxwell model given in Eq.~\eqref{eq:general_jeffreys}. The elements in the circuit correspond to four-tensors and adhere to odd viscoelastic Kirchhoff laws \cite{Banerjee_2021,lier2021passive}. They are related to the tensors in Eq.~\eqref{eq:general_jeffreys} by $
         \ull{\alpha}  =   ( \ull{C_1} : \ull{K^{-1}_1} + \ull{C_2} : \ull{K^{-1}_2} )$, $\ull{\beta}  = \ull{C_1} : \ull{C_2}  :\ull{K_1}^{-1}  :\ull{K_2}^{-1}$, $\ull{\sigma} = ( \ull{C_1} +  \ull{C_2} )$ and $\ull{\xi} =  (\ull{C_1} :\ull{C_2} :\ull{K^{-1}_1}+\ull{C_1}: \ull{C_2} : \ull{K^{-1}_2})$, where the colon represents a contraction of the last two indices of one tensor with the first two indices of another: $(\ull{\alpha}:\ull{\beta})_{ijkl} = \alpha_{ijmn}\beta_{mnkl}$.}
            \label{fig:circuit}
\end{figure}

The non-equilibrium part $\tilde{T}_{ij}$ evolves in time in a non-trivial way due to the behavior of the component $T^\mathrm{(\mathrm{s})}_{ij}$. Its evolution can be found by multiplying the Fokker-Planck equation in Eq. (\ref{fp_eq}) by $R_jF_i(\vct R)$ and integrating over $\vct R$, as implied by Eq. (\ref{eq:Ts}). The resulting equations are then simplified by expanding up to the first order in gradients, as well as to the lowest order in $\langle\delta R^2\rangle/R^2$. The details of this lengthy calculation can be found in Appendix ~\ref{app:modelderivation}. The main finding is that the evolution of $\tilde{T}_{ij}$ is described by a generalized odd Maxwell model, the equation for which involves second time derivatives. In a symbolic notation we can write 
\begin{equation}
    \frac{\partial^2}{\partial t^2}\tilde{\underline T} + \underline{\underline{\alpha}} \frac{\partial}{\partial t}\tilde{\underline T} + \underline{\underline{\beta}} \tilde{\underline T} = -  \underline{\underline{\sigma}} \frac{\partial}{\partial t}\underline D -  \underline{\underline{\xi}} \underline D\,,
    \label{eq:general_jeffreys}
\end{equation}
where all the symbols with underlines denote tensors: namely, $\tilde{\underline T}=\tilde{T}_{ij}$ and $\underline D=D_{ij}$ are rank-two tensors, while $\underline{\underline{\alpha}}=\alpha_{ijkl}$, $\underline{\underline{\beta}}=\beta_{ijkl}$, $\underline{\underline{\sigma}} = \sigma_{ijkl}$ and $\underline{\underline{\xi}} = \xi_{ijkl}$ are rank-four tensors. A schematic representation of Eq. (\ref{eq:general_jeffreys}) in the form of a viscoelastic circuit is shown in Fig. \ref{fig:circuit}. All the rank-four tensors in Eq. (\ref{eq:general_jeffreys}) can be analytically calculated, however, their physical meaning is not immediately clear in this formulation. In order to elucidate it, we notice that when the frequency of the perturbation is sufficiently high, the third term on the left hand side and the second term on the right-hand side can be neglected; conversely, when the frequency is low enough, the first term on the left-hand side can be neglected. Therefore, if the timescales in the problem are well separated, the generalized Maxwell model reduces to two odd Jeffreys models, one describing the high-frequency evolution, and one describing the low-frequency evolution of stress. Splitting the stress tensor $\tilde{T}_{ij}$ into its traceless symmetric component, trace, and the anti-symmetric component, we then obtain the following set of equations valid for both regimes:
\begin{subequations}
\begin{align}
\begin{split} \label{eq:shearsector}
    \frac{\partial}{\partial t } \tilde{T}_{\langle  i j  \rangle } 
      +   \chi_s  \tilde{T}_{\langle ij  \rangle } &   + \chi_{o}  \tilde{T}^o_{\langle ij  \rangle } = \\
    =  - \gamma_s \frac{\partial}{\partial t } D_{\langle  i j  \rangle }& -   \left(\zeta_s +\gamma_s\chi_s-\gamma_o\chi_o \right)  D_{\langle ij \rangle }   \\
    - \gamma_o \frac{\partial}{\partial t } D^o_{\langle  i j  \rangle }&-   \left(\zeta_{o} +\gamma_o\chi_s+\gamma_s\chi_o\right)    D^o_{\langle ij \rangle }\,, 
\end{split} \\
     \frac{\partial}{\partial t }\tilde{T}  + \chi_b \tilde{T}   =  -\gamma_b& \frac{\partial}{\partial t } D  - \left(\zeta_b+\gamma_b\chi_b\right) D\,,
\end{align} \label{eq:passiveviscogeneralmodel}
\end{subequations}
while the antisymmetric component is related to the trace of $\tilde{T}_{ij}$:
\begin{equation}
    \tilde{T}^A = \frac{\kappa_o}{\kappa_e}\left(\tilde{T}+\eta_b D\right)\,.
\end{equation}
Equations (\ref{eq:passiveviscogeneralmodel}) represent the odd Jeffreys model of viscoelasticity.
The interpretation of the different coefficients is as follows: $\chi_s$ gives the stress relaxation rate; $\chi_o$ gives the frequency of rotation of stress in the $\tilde{T}_{\langle xx\rangle}-\tilde{T}_{\langle xy\rangle}$ space; $\gamma_s$ and $\gamma_o$ are the instantaneous shear and odd viscosities, respectively; and $\zeta_s$ and $\zeta_o$ are the instantaneous shear and odd elasticities, respectively. Furthermore, note that for times $t\gg \chi_s^{-1}$, after the relaxation has taken place, the stress tends to
\begin{subequations}
\begin{align}
    \tilde{T} \rightarrow & -\left(\frac{\zeta_b}{\chi_b}+\gamma_b\right)D\,, \\
    \tilde{T}^A \rightarrow & -\frac{\kappa_o}{\kappa_e}\left(\frac{\zeta_b}{\chi_b}+\gamma_b-\eta_b\right)D\,, \\
    \begin{split}
    \tilde{T}_{\langle ij\rangle} \rightarrow & -\left(\frac{\chi_s\zeta_s+\chi_o\zeta_o}{\chi_s^2+\chi_o^2}+\gamma_s\right)D_{\langle ij\rangle} \\
    & -\left(\frac{\chi_s\zeta_o-\chi_o\zeta_s}{\chi_s^2+\chi_o^2}+\gamma_o\right)D^o_{\langle ij\rangle} \label{eq:long_viscosities_shear}\,.
    \end{split}
\end{align} \label{eq:long_viscosities}
\end{subequations}

In the following we consider odd dumbbell models with $\kappa_e \gg \zeta \Omega$, which is a valid approximation in systems in which the relaxation of elastic deformations takes place much faster than the rotation period. In this case we can distinguish two regimes of interest. One corresponds to the evolution of the system at times on the order of $\zeta/\kappa_e$ or shorter: this is the timescale on which the relaxation of the elastic strain given by
\begin{align}
    C_{ij}  =  \int_{\vct R}\hat{r}_i \hat{r}_j\delta R\,\psi 
\end{align}
takes place. The coefficients in this regime are as follows:
\begin{subequations}
\begin{align} 
    \gamma_s & = \eta_s\,, \quad \quad \chi_s= \frac{2\kappa_e}{\zeta}\,, \quad\quad \zeta_s = \frac{n^\mathrm{d}R_*^2\kappa_e}{4}\,, \\
    \gamma_o & = 0\,, \quad\quad \chi_o=0\,,\quad\quad \zeta_o = \frac{n^\mathrm{d}R_*^2\kappa_o}{4}\,, \\
    \gamma_b & = \eta_b\,,\quad\quad \chi_b = \frac{2\kappa_e}{\zeta}\,,\quad\quad \zeta_b= \frac{n^\mathrm{d}R_*^2\kappa_e}{2}\,.
\end{align} \label{eq:coeffs_1}
\end{subequations}
The second regime corresponds to times on the order of $\zeta/\kappa_e$ or longer: it is on this timescale that the alignment tensor $X_{ij}$ relaxes, with $X_{ij}$ defined as
\begin{align}
    X_{ij} \equiv R_*^2 \int_{\vct R} \hat{r}_i\hat{r}_j\psi \, . 
\end{align}
The corresponding coefficients are
\begin{subequations}
    \begin{align}
    \begin{split}
    \gamma_s & = \eta_s+\frac{n^\mathrm{d}R_*^2\zeta}{8}\,, \quad\quad\chi_s = \frac{8kT}{\zeta R_*^2}\left(1+\frac{\kappa_o^2}{\kappa_e^2}\right)\,,\\
    \zeta_s&=\frac{n^\mathrm{d}R_*^2\zeta}{8}\left(\chi_s+\frac{\kappa_o}{\kappa_e}\chi_o\right)\,, 
    \end{split}
    \\ \begin{split}
    \gamma_o & = \frac{\kappa_o}{\kappa_e}\frac{n^\mathrm{d}R_*^2\zeta}{8}\,,\quad \chi_o = 2\Omega-\frac{4kT \kappa_o}{\kappa_e R_*^2\zeta}\,,\\
    \zeta_o & = \frac{n^\mathrm{d}R_*^2\zeta}{8}\left(\chi_o-\frac{\kappa_o}{\kappa_e}\chi_s\right)\,,\end{split}\\
    \gamma_b &= \eta_b + \frac{n^\mathrm{d}R_*^2\zeta}{4}\,, \quad \chi_b=0\,, \quad \zeta_b=0\,.
\end{align} \label{eq:coeffs_2}
\end{subequations}
Some interesting observations can be made at this point with the help of Eq. (\ref{eq:long_viscosities}). Firstly, the post-relaxation viscosities in the short-time regime that can be read off from Eqs. (\ref{eq:long_viscosities}) and (\ref{eq:coeffs_1}) are equal to the instantaneous viscosities $\gamma_b$, $\gamma_s$ and $\gamma_o$ in the long-time regime in Eq. (\ref{eq:coeffs_2}). This agreement can be seen as a check on the consistency of the two-regime approach. Secondly, for $t\rightarrow\infty$ the stress tensor tends to
\begin{subequations}
\begin{align}
    \tilde{T} \rightarrow & -\left(\eta_b + \frac{n^\mathrm{d}R_*^2\zeta}{4}\right)D\,, \\
    \tilde{T}^A \rightarrow & -\frac{\kappa_o}{\kappa_e} \frac{n^\mathrm{d}R_*^2\zeta}{4}D\,, \\
    \tilde{T}_{\langle ij\rangle} \rightarrow & -\left(\eta_s + \frac{n^\mathrm{d}R_*^2\zeta}{4}\right)D_{\langle ij\rangle}\,.
\end{align} \label{eq:long_viscosities_ours}
\end{subequations}
Clearly, odd viscosity vanishes in the zero-frequency limit. While it might seem to be a nontrivial result given the complicated expression in Eq. (\ref{eq:long_viscosities_shear}), we will show that it can be expected on general grounds. To this end, let us consider the evolution of the object
\begin{equation}
    U_{ij} \equiv \int_{\mathbf{R}} R_i R_j\psi \,  . 
\end{equation}
Multiplying the Fokker-Planck equation in Eq.~(\ref{fp_eq}) by $R_iR_j$ and calculating the average produces the evolution equation for $U_{ij}$:
\begin{multline}
    \frac{\mathcal{D}}{\mathcal{D} t}U_{ij} +U_{ij} \partial_k u_k  = 2U_{(ik}\partial_k u_{j)}+\frac{4}{\zeta}T^{(s)}_{(ij)}+\\
    +\frac{4kT}{\zeta} \delta_{ij}+\frac{n^{\mathrm{d}}kT}{4\zeta}\Delta_x U_{ij}\,,
\end{multline}
where $\frac{\mathcal{D}}{\mathcal{D} t} \equiv \frac{\partial}{\partial t} + \vct u \cdot \nabla$. Let us assume that after a constant shear rate is turned on, $U_{ij}$ first builds up and then relaxes to a steady value, so that at long times $\frac{\mathcal{D}}{\mathcal{D} t} U_{ij} \rightarrow 0$. Due to the isotropy of the system, in the hydrodynamic regime $U_{ij} =  U_s  \delta_{ij}+\mathcal{O}(\partial)$,  where $U_s$ is some constant and $\mathcal{O}(\partial)$ is a gradient correction. We thus have at the leading order 
\begin{equation}
    T^{(s)}_{(ij)} \rightarrow -n^{\mathrm{d}}kT\delta_{ij}-\frac{\zeta}{2 } 
U_s \partial_{\langle i} u_{j\rangle}\,.
\end{equation}
Thus, in the $t \rightarrow \infty$ limit there is only the shear viscosity induced by drag. This conclusion is reached assuming only (1) an overdamped collisionless dumbbell model, in which (2) the tensor $U_{ij}$ reaches a steady state. For our model, $U_s$ is given by
\begin{equation}
 U_s  =    \frac{n^{\mathrm{d}} R_*^2}{2}  \,,
\end{equation}
and $U_{ij}$ relaxes to $U_{ij} =  U_s  \delta_{ij}+\mathcal{O}(\partial)$ whenever the temperature is nonzero. This reproduces the results of Eq.~(\ref{eq:long_viscosities_ours}).

On the other hand, we also observe that the stress relaxation rate is proportional to $kT$, so that in the zero-temperature limit odd viscosity will be equal to its instantaneous value $\gamma_o = \frac{\kappa_o}{\kappa_e}\frac{n^\mathrm{d}R_*^2\zeta}{8}$ even for $t\rightarrow\infty$. We confirm these predictions in numerical simulations presented in the next section.

\section{Simulation results}
\label{sec:numerical}

\begin{figure*}
    \subfloat{
        \includegraphics[width=0.48\linewidth]{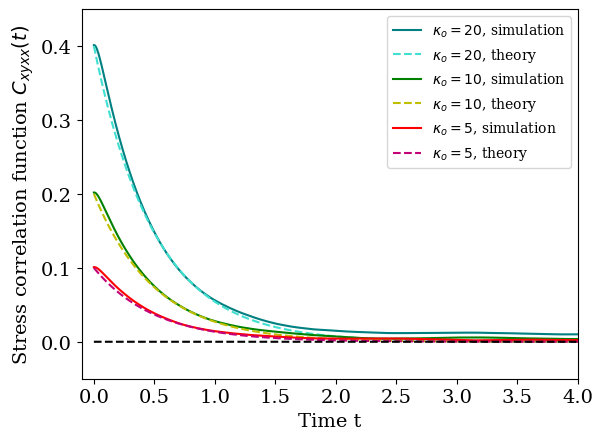}
    }\hfill
    \subfloat{
        \includegraphics[width=0.485\linewidth]{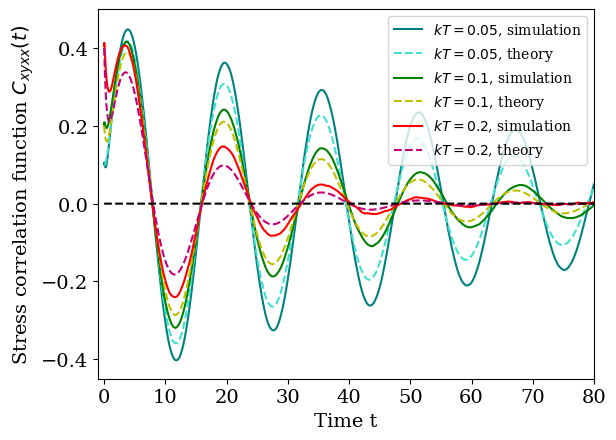}
    }\hfill
\caption{Comparison of numerical results for the correlation functions $C_{xyxx}(t)=\langle \tilde{T}_{xy}(t) \tilde{T}_{xx}(0)\rangle$ with the analytical result of Eq.~(\ref{eq:Cxyxx}). In the left panel we present results for different odd spring constants $\kappa_o$ in the short-time regime, setting $\zeta\Omega=0$ and $kT=0.2$. In the right panel we present results for different temperatures $kT$ across both regimes, setting $\zeta\Omega=6$ and $\kappa_o=20$. In all simulations $\kappa_e = 30$, $\zeta=30$, $n^\mathrm{d}=0.4$, $R_*^2=1$.}
\label{fig:plots1}
\end{figure*}

To show that the odd viscoelastic models discussed in this paper can indeed describe a system composed of a large amount of particles described by microscopic equations, we perform numerical simulations for a many-particle system where we trace the evolution of the stress tensor. The goal of the simulations is to test the accuracy of the coarse-graining analysis, verifying the validity of the approximations that were employed in obtaining the analytical results. Furthermore, numerical studies give an opportunity to study the effects of inter-dumbbell interactions on the odd viscoelastic behavior. As the central result of this work consists of predicting the evolution of stress under applied shear, see Eq.~\eqref{eq:general_jeffreys}, we concentrate on that part of the problem by subjecting the system to a constant shear rate while ignoring the dynamics of the background fluid itself.

The computations are based on the publicly available code developed in Ref. \cite{epsteinnemd}, to which we introduced adequate modifications. The simulation consists of two parts. First, we consider exclusively the evolution of the stress in absence of shear and we find that in the presence of inter-dumbbell interactions the qualitative behavior that is predicted by the analytics is retained. Then, we consider a model in the presence of shear, which will enable us to study the full viscoelastic response. More details about the simulations can be found in the subsequent subsections.

\subsection{Stress tensor evolution without shear}
We consider a setup where the background fluid is fully homogeneous and at rest, so that $D_{ij}=0$. The evolution of stress is then given by
\begin{equation}
    \partial_t \begin{pmatrix}
        \tilde{T}_{\langle xx \rangle} \\
        \tilde{T}_{\langle xy \rangle} \\
        \frac12\tilde{T} \\
        \frac12\tilde{T}^A
    \end{pmatrix} = 
    M \begin{pmatrix}
        \tilde{T}_{\langle xx \rangle} \\
        \tilde{T}_{\langle xy \rangle} \\
        \frac12\tilde{T} \\
        \frac12\tilde{T}^A
    \end{pmatrix}\,,
\end{equation}
with 
\begin{equation}
    M = \begin{pmatrix}
     -\chi_s & -\chi_o & 0 &0\\
     \chi_o & -\chi_s & 0 &0\\
     0&0&-\chi_b&0 \\
      0&0&0&-\chi_b
    \end{pmatrix}\,.
\end{equation}
The viscoelastic coefficients in this setup can be found numerically by measuring stress-stress correlation functions. This can be done based on the code developed in Ref. \cite{epsteinnemd}, in which we modify the formula for the intra-dumbbell force. In order to simulate the odd-dumbbell model introduced in Sec. \ref{sec:springpart}, we take the force to be
\begin{equation}
    F_i = \delta R\,\kappa_e \hat{r}_i + \varepsilon_{ij}\hat{r}_j \frac{R_*}{R}\left(\kappa_o  \delta R+\frac{\zeta\Omega}{2} R_*\right)\,.
\end{equation}
A system of dumbbells is then subjected to a Langevin evolution and allowed to thermalize, after which the stress-stress correlator is measured during a longer period of evolution.

We shall focus by the way of example on the correlation function $C_{xyxx}(t)=\langle \tilde{T}_{xy}(t) \tilde{T}_{xx}(0)\rangle$, the formula for which reads
\begin{equation}
      C_{xyxx}(t) = \langle \tilde{T}_{\langle xx \rangle}^2\rangle e^{-\chi_s t}\sin(\chi_o t) + \frac{1}{4}\langle \tilde{T}\cdot\tilde{T}^A\rangle e^{-\chi_b t}
\end{equation}
with $\langle \ldots \rangle$ denoting an equal-time correlator.
Using $\langle (\delta R)^2 \rangle = kT/\kappa_e$ and isotropy of the distribution of $\hat{r}_i$ in a steady state one can prove
\begin{equation}
\begin{split}
    \langle \tilde{T}_{\langle xx \rangle}^2\rangle & = \frac{n^\mathrm{d}  R_*^2 kT  \kappa_e}{8}\left(1+\frac{\kappa_o^2}{\kappa_e^2}\right) \\ & \quad + \frac{n^\mathrm{d} R_*^4 \zeta^2\Omega^2}{32} +\mathcal{O}\left((kT)^2\right)\,, \\
    \langle \tilde{T} ^2\rangle & =  \kappa_en^\mathrm{d}  R_*^2 kT, \quad \langle \tilde{T} \cdot \tilde{T}^A\rangle = \kappa_on^\mathrm{d}  R_*^2 kT\,.
\end{split}
\end{equation}
The values given above are only valid for the short-time evolution, after which a relaxation of the elastic part of the stress tensor takes place. Calculating the post-relaxation correlation functions requires some care, the appropriate values are calculated in Appendix \ref{app:modelderivation} in Eq. (\ref{eq:stress_corr_relaxed}). One finds that in the long-time regime $\langle \tilde{T}^2\rangle = \langle \tilde{T} \cdot \tilde{T}^A\rangle= 0$, while for $\langle \tilde{T}_{\langle xx \rangle}^2\rangle$ the following replacement should be made: 
\begin{equation}
    \langle \tilde{T}_{\langle xx \rangle}^2\rangle \rightarrow \frac{n^\mathrm{d} R_*^4 \zeta\Omega}{32}\left(\zeta\Omega-\frac{4kT\kappa_o}{\kappa_e R_*^2}\right) +\mathcal{O}\left((kT)^2\right).
    \label{eq:corrs_replace}
\end{equation}
Finally, the formula that captures both the short-time and long-time behavior is
\begin{multline}
    C_{xyxx}(t) = \frac{n^\mathrm{d} \kappa_o R_*^2kT}{4}\exp\left(-\frac{2\kappa_e}{\zeta}t\right) \\
    +  \frac{n^\mathrm{d}R_*^4\zeta^2\Omega_*^2}{32}\exp\left[-\frac{8kT}{\zeta R_*^2}\left(1+\frac{\kappa_o^2}{\kappa_e^2}\right)t\right]\sin\left(2\Omega t\right)\,,
    \label{eq:Cxyxx}
\end{multline}
where we assumed for simplicity that $\zeta\Omega \gg kT\kappa_o/(\kappa_e R_*^2)$ in the second line of the equation. In Fig. \ref{fig:plots1} the behavior of $C_{xyxx}(t)$ obtained by the means of numerical simulations is compared to the analytically-derived formula given in Eq.~\eqref{eq:Cxyxx}. While the approximations used in the derivation are seen to introduce some discrepancies between the two results, the overall behavior turns out to be accurately described by the theory, especially regarding the scaling of the various coefficients with the parameters of the system. 

\begin{figure*}
    \subfloat{
        \includegraphics[width=0.225\linewidth]{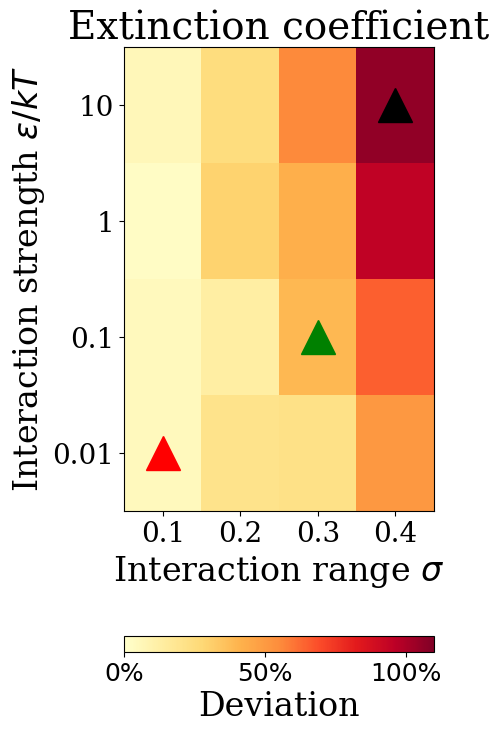}
    }\hfill
    \subfloat{
        \includegraphics[width=0.22\linewidth]{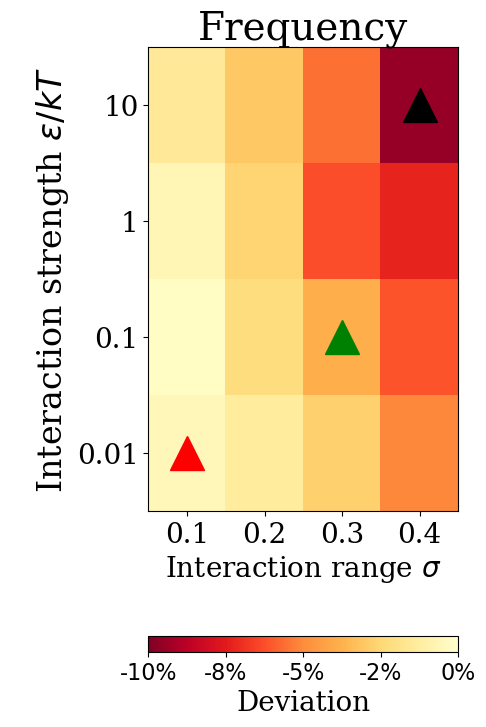}
    }\hfill
    \subfloat{
        \includegraphics[width=0.48\linewidth]{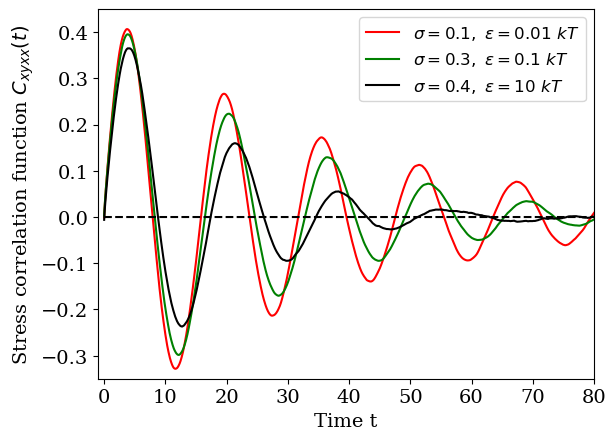}
    }\hfill
\caption{Effect of introducing bead-bead interactions modelled by the Lennard-Jones potential with range $\sigma$ and strength $\epsilon$. (a) A function of the form $A\exp(-\gamma t)\cos(2\Omega_*t)$ is fitted to the numerically obtained correlation function $C_{xyxx}(t)$. The diagrams present the relative deviation of the extinction coefficient $\gamma$ and the frequency $\Omega_*$ from their interaction-free values $\gamma_0$ and $\Omega$; the deviations are defined as $\frac{\gamma - \gamma_0}{\gamma_0}$ and $\frac{\Omega_* - \Omega}{\Omega}$ respectively. (b) Three examples of the correlation functions $C_{xyxx}(t)$ obtained for different values of the interaction parameters. The plots correspond to the data points from panel (a) marked by triangles with colors (shades) matching the plots. The parameters used are $\kappa_e = 30$, $\kappa_o=0$, $\zeta=30$, $\zeta\Omega=6$, $n^\mathrm{d}=0.4$, $R_*^2=1$ and $kT=0.1$.}
\label{fig:interactions}
\end{figure*}

Furthermore, to test the limits of applicability of the model we introduced bead-bead interactions in the form of the Lennard-Jones potential:
\begin{equation}
    E = 4 \epsilon \sum_{i<j}\left[ \left(\frac{\sigma}{r_{ij}}\right)^{12} -
        \left(\frac{\sigma}{r_{ij}}\right)^6 \right],
\end{equation}
where $r_{ij}=|\vct r_i- \vct r_j|$ is the distance between beads numbered $i$ and $j$. We focused on the case $2\sigma < R_*$, which means that the bead-bead interactions happen mostly during inter-dumbbell collisions, while intra-dumbbell collisions are rare. As seen in Fig.~\ref{fig:interactions} interactions are found to increase the oscillation period while reducing the relaxation time of the out-of-equilibrium stress tensor, but the overall shape of the stress correlation function remains broadly intact. This behavior can be observed across a wide range of interaction parameters, with the relative change of the oscillation frequency and extinction coefficients plotted in Fig.~\ref{fig:interactions}. Noticeably, the increase of the extinction coefficient is typically a more significant effect than the decrease of the frequency.

\subsection{Stress tensor evolution with shear}

\begin{figure*}
    \subfloat{
        \includegraphics[width=0.485\linewidth]{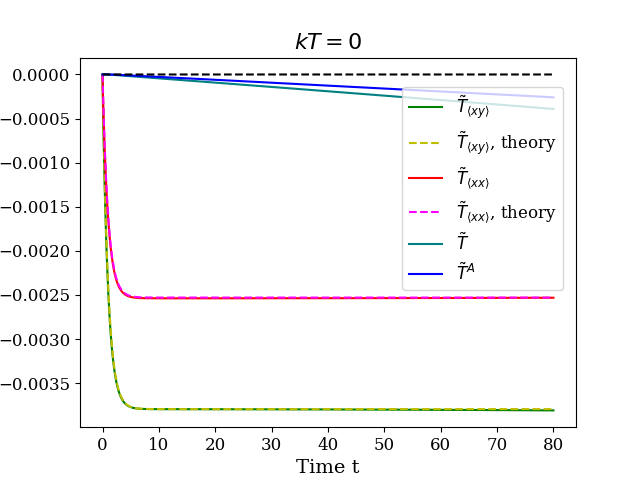}
    }\hfill
    \subfloat{
        \includegraphics[width=0.485\linewidth]{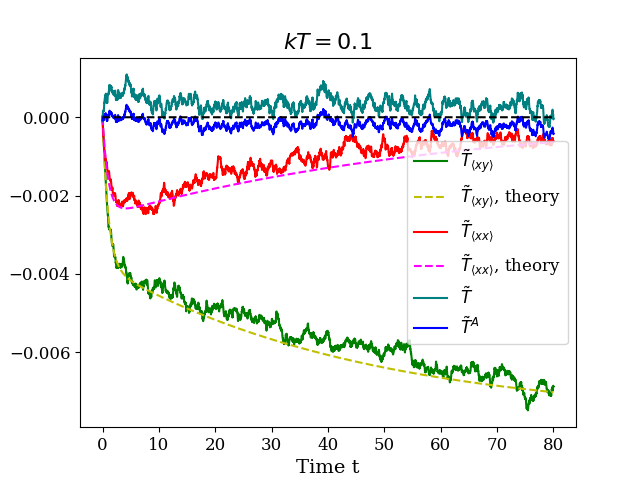}
    }\hfill
\caption{Evolution of all the components of stress under a constant simple shear $D_{xy}=0.00126$. Numerical results (solid lines) are compared with analytical predictions (dashed lines). The $\tilde{T}_{\langle xy \rangle}$ and $\tilde{T}_{\langle xx\rangle}$ components of stress corresponds to the even and odd responses, respectively; for the trace and antisymmetric components, the Jeffreys model predicts $\tilde{T}=\tilde{T}^A=0$. The left panel shows the zero-temperature case $kT=0$, while in the right panel $kT=0.1$. In both simulations we set $\Omega=0$, $\kappa_e = 30$, $\kappa_o=20$, $\zeta=30$, $n^\mathrm{d}=0.4$, $R_*^2=1$. The oscillations visible in the right panel result from finite-temperature fluctuations coupled with a finite size of the numerical sample.}
\label{fig:plots3}
\end{figure*}

Now we turn to a simulation that allows us to directly test the validity of the viscoelastic model as given in Eqs. (\ref{eq:passiveviscogeneralmodel}), (\ref{eq:coeffs_1}) and (\ref{eq:coeffs_2}). While using a similar setup to the one described in the previous subsection, we now impose a small constant deformation rate on the region of simulation starting at a certain time $t=0$. The deformation imposed is in the form of a simple shear equivalent to imposing $D_{\langle xy \rangle} = \frac12 D^A = \mathrm{const}.$ and $D_{\langle xx\rangle}=D=0$. Then the average stress tensor of the system of dumbbells is directly measured. As we do not model the dynamics of the background fluid, we set $\eta_s=\eta_o=\eta_b=0$. The results are shown in in Fig. \ref{fig:plots3}, where they are compared with the analytical prediction that combines both the short-time and the long-time effects: 
\begin{equation}
    \tilde{T}_{\langle ij\rangle} = -\frac{n^\mathrm{d}R_*^2\zeta}{8}\left(A(t)D_{\langle ij\rangle}+B(t)D_{\langle ij\rangle}^o\right)\,,
\end{equation}
where 
\begin{subequations}
    \begin{align}
    \begin{split}
        A(t) =& \left(1-e^{-2\kappa_e t/\zeta}\right)\left(2-e^{-\chi_s t}\right)\\
        &\times\left(\cos\left(\chi_o t\right)+\frac{\kappa_o}{\kappa_e}\sin\left(\chi_o t\right)\right)\,,\end{split}\\
        \begin{split}
        B(t) =& \frac{\kappa_o}{\kappa_e}\left(1-e^{-2\kappa_e t/\zeta}\right)\left(2-e^{-\chi_s t}\right)\\
        &\times \left(\cos\left(\chi_o t\right)-\frac{\kappa_e}{\kappa_o}\sin\left(\chi_o t\right)\right)\,,
        \end{split}
    \end{align}
\end{subequations}
where the coefficients $\chi_s$ and $\chi_o$ relate to the long-time regime and are found in Eq. (\ref{eq:coeffs_2}). At the same time the trace and the antisymmetric parts of stress are predicted to be zero, $\tilde{T}=\tilde{T}^A=0$. 

The simulations fully corroborate the results of Sec. \ref{sec:jeffreys}, apart from the fact that $\tilde{T}$ and $\tilde{T}^A$ are found to grow slowly with time; this behavior can be attributed to effects non-linear in $D_{ij}$, which are not captured by the linear Jeffreys model. In particular, the simulations confirm that the long-term odd viscosity tends to zero as a result of thermal fluctuations, but remains constant in the zero temperature case.

\section{Conclusion}
In this work we consider a microscopic model of dumbells with odd active forces suspended in a fluid. We provide two examples of systems in which such a suspension of odd dumbbells considered could appear, focusing particularly on the relation between these odd forces and angular momentum conservation.
\newline 
In the overdamped limit, we coarse-grain the microscopic equations of these systems to obtain a macroscopic description of the dumbbell suspension, which is the theory of odd viscoelasticity. Specifically, we find two distinct regimes in time where two distinct objects relax to their equilibrium value; both of these regimes can be described by their own distinct odd Jeffreys model. One of these regimes, lasting for a time of the order of $\zeta/\kappa_e$, where $\zeta$ is the drag coefficient and $\kappa_e$ is the elastic constant of a single colloidal particle, can be seen as the elastic regime: it is at this timescale that odd-elastic effects can be observed. Following it is a time window $\zeta/\kappa_e \ll t \ll R_*^2 \zeta/(kT)$, where $R_*$ is the radius of the particle and $kT/\zeta$ is the effective diffusion constant, when the system displays odd viscosity. Finally, at times of the order of $R_*^2 \zeta/(kT)$ the odd response dies out. We have also shown that the vanishing of odd response at long times is a feature that can generically be expected in this class of systems, which, however, does not prevent the emergence of transient (or, equivalently, finite-frequency) odd elastic and odd viscous behaviors, as proven by our analysis.

Through numerical simulation, we verify that the coarse-graining procedure provides quantitatively accurate predictions for time-dependent correlations of the elastic stress tensor, including in the presence of weak interdumbbell interactions.
\newline 
The analytical and numerical results of this work offer a foundation for potentially observing odd viscoelastic fluids in experiment. Furthermore, the coarse-graining procedure provides a deeper understanding of the circumstances under which the coefficients of an odd viscoelastic fluids can arise.

\section{Acknowledgments}
 P.M. was supported by the Deutsche Forschungsgemeinschaft (DFG) through the cluster of excellence ct.qmat (EXC 2147, project-id 390858490). P.S. acknowledges support from the Polish National Science Centre (NCN) Sonata Bis grant 2019/34/E/ST3/00405. We thank Yuchao Chen and Nikta Fakhri for discussions and comments on the manuscript.

\addcontentsline{toc}{chapter}{Bibliography}
\bibliography{biblio}

\onecolumngrid

\appendix

\section{Odd spring model}
\label{app:oddspring}

\begin{figure}[h]
    \includegraphics[width=0.4\linewidth]{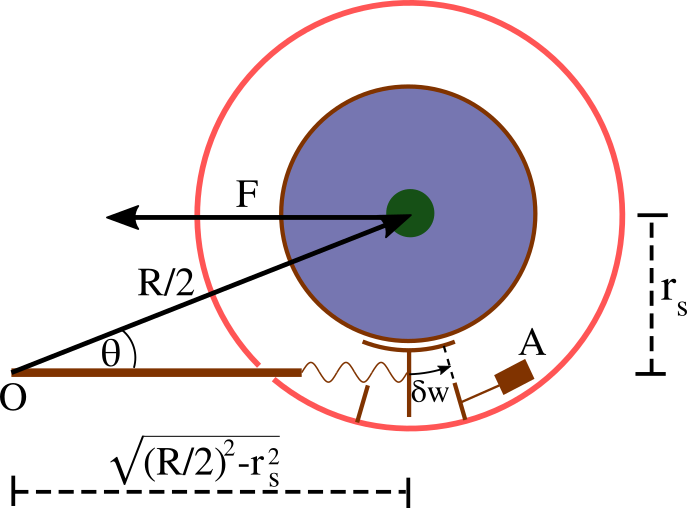}
    \caption{Schematic representation of one half of the odd dumbbell from Fig.~\ref{fig:torque_model} as seen from above, with $O$ denoting the center of the dumbbell. A more detailed description of the Figure is the content of Appendix \ref{app:oddspring}.}
            \label{fig:odd_spring}
\end{figure}

In this appendix we present a possible mechanism of generating odd spring force acting on the dumbbell. The diagram in Fig.~\ref{fig:odd_spring} should be read as follows: there is a rigid rod connecting the two beads constituting the dumbbell (the thick brown line in the sketch). Due to symmetry between the two beads we concentrate on just one half of the dumbbell. The vector $\vct R/2 = \frac{R_*+\delta R}{2}\hat{\vct r}$ connects the center of the dumbbell with the center of the bead and we denote the radius of the cylinder by $r_s$. We also introduce a coordinate $z$ equal to the displacement of the bead \textit{along the direction of the rod} from its rest state. In other words, 
\begin{equation}
    z = \sqrt{\frac{(R_*+\delta R)^2}{4}-r_s^2}-\sqrt{\frac{R_*^2}{4}-r_s^2}\,.
\end{equation}
The end of the rod is attached to a spring, which in turn exerts force on the cylinder through some mechanism. If the other end of the spring deviates from its original position \textit{with respect to the bead} by $w$, the elastic energy of the spring is $\frac12 \kappa (z+w)^2$ with some spring constant $\kappa$. The spring exerts a force at the cylinder, which is applied at the point of contact with the cylinder and directed along the rod. Thus, the cylinder experiences linear acceleration due to the force $F = -\kappa (z+w)$ (shown on the sketch) and angular acceleration due to the torque $M=r_s F$.

Let us now turn to energetic considerations. To rotate the cylinder, the spring needs to do work on it: to be exact, the work done when the end of the spring is displaced by $w$ is equal to $\frac12\kappa  \,w^2$. On the other hand, the internal degree of freedom $w$ does not feature in the equations of motion for the odd dumbbell. One possibility for eliminating $w$ from the dynamics of the dumbbell is by keeping $w$ roughly equal to $0$ all the time by using a device (denoted $A$ on the sketch) that pushes the end of the spring back to $w=0$ when $w$ reaches a certain value, which we denote as $\pm\delta w$. This device is necessarily active, as every activation of the mechanism costs energy $\frac12\kappa (\delta w)^2$, and so requires a source of energy to run.

Having assumed now the presence of the active device $A$ thanks to which always $w\approx 0$, let us calculate the forces acting on the bead. The force $F$ can be projected along and perpendicular to $\vct R$, and in the situation shown in the sketch it is equal to
\begin{equation}
\begin{split}
	F_i  &= -\kappa z \left( \cos(\theta )
 \hat{r}_i+ 
\sin(\theta ) \varepsilon_{ij}\hat{r}_j\right) \\  
 &= -\kappa z \left(\sqrt{1-\frac{4 r_s^2}{(R_*+\delta R)^2}}\hat{r}_i+\frac{2r_s}{R_*+\delta R}\varepsilon_{ij}\hat{r}_j\right) \\
	 &= - \frac{\kappa}{2}\left(\sqrt{(R_*+\delta R)^2-4r_s^2}-\sqrt{R_*^2-4r_s^2}\right) \left(\sqrt{1-\frac{4r_s^2}{(R_*+\delta R)^2}}\hat{r}_i+\frac{2r_s}{R_*+\delta R}\varepsilon_{ij}\hat{r}_j\right)\\
	& \approx  -\kappa \frac{R_*}{\sqrt{R_*^2-4r_s^2}} \left(\frac{\sqrt{R_*^2-4r_s^2}}{R_*}\hat{r}_i+\frac{2r_s}{R_*+\delta R}\varepsilon_{ij}\hat{r}_j\right)\delta R\,,
\end{split}
\end{equation}
which can be expressed in the form 
\begin{equation}
	F_i \approx -\left(\kappa_e\hat{r}_i+\frac{R_*}{R_*+\delta R}\kappa_o\varepsilon_{ij}\hat{r}_j\right)\delta R\,.
\end{equation}

\section{Derivation of the stress tensor}
\label{app:stressderivation}

In this appendix we derive the equation for momentum conservation which includes the stress tensor of Eq.~\eqref{eq:Ts}. In the main text we formulated the microscopic equations for the overdamped limit. To study momentum conservation, we must first extend these equations for the dumbbell components given by Eq.~\eqref{eq:stochasticstart} to include inertia, to wit
\begin{align} 
     m\ddot{\vct r}^{(\lambda)} + \zeta\left(\dot{\vct r}^{(\lambda)}-\vct u^\mathrm{f}(\vct r^{(\lambda)})\right)+(-1)^{\lambda} \vct F(\vct R) - \vct S^{(\lambda)} & = 0\,,
     \label{eom_original1}    
\end{align}
where $\lambda \in \{\alpha,\beta\} $ and 
\begin{align}
    (-1)^{\{\alpha,\beta\}} = \{\ -1,1\}\,. 
\end{align}
Eq.~\eqref{eom_original1} can be rewritten to 
\begin{subequations}
\begin{align}
    \dot{\vct r}^{(\lambda)} &= \vct v^{(\lambda)}, \\     \dot{\vct v}^{(\lambda)} &= -\frac{\zeta}{m}\left(\vct v^{(\lambda)}- \vct u^\mathrm{f} (\vct r^{(\lambda)}) \right)- (-1)^{\lambda} \frac{1}{m}\vct F(\vct R)+\frac{1}{m}\vct S^{(\lambda)}\,. 
 \end{align}    \end{subequations}
We can then obtain the Fokker-Planck equation \cite{tong2012lectures}
   \begin{align} 
\begin{split}
    \left(\frac{\partial}{\partial t}+ 
\sum_{\lambda } \vct v^{(\lambda)}\cdot \nabla_{r^{(\lambda)}}\right)\psi =  \sum_{\lambda } \left\{  \frac{\zeta}{m}\nabla_{v^{(\lambda)}} \cdot  \left[(\vct v^{(\lambda)} -   \vct u^\mathrm{f} (\vct r^{(\lambda)}) )\psi\right]      + (-1)^\lambda \frac{1}{m}\vct F(\vct r^{(\beta)}-\vct r^{(\alpha)}) \cdot \nabla_{v^{(\lambda)}}\psi + \frac{kT \zeta}{ m^2}\Delta_{v^{(\lambda)}} \psi  \right\} \,.
\label{fp_x1x2} 
\end{split}
   \end{align}
Let us define a local average of a quantity $A$ over the distributions of beads:
\begin{align}
    \langle A \rangle^{(\lambda)} \equiv \int_{\vct r^{(\alpha)}, \vct r^{(\beta)}, \vct v^{(\alpha)}, \vct v^{(\beta)}} A (\vct r^{(\alpha)}, \vct r^{(\beta)}, \vct v^{(\alpha)}, \vct v^{(\beta)}, t) \psi \delta(\vct r - \vct r^{(\lambda)})\,, 
\end{align}
where we employ the notation $\int_{\vct r^{(\alpha)}, \vct r^{(\beta)}, \vct v^{(\alpha)}, \vct v^{(\beta)}} \equiv \int d^2 r^{(\alpha)}d^2 r^{(\beta)}d^2 v^{(\alpha)}d^2 v^{(\beta)}$. Let us define $n^{(\lambda)}\equiv \langle 1 \rangle^{(\lambda)}$ and $\bar{\vct v}^{(\lambda)}\equiv \langle \vct v^{(\lambda)}\rangle^{(\lambda)}/ n^{(\lambda)}$. A simple integration of Eq. \eqref{fp_x1x2} gives
\begin{align} \label{eq:densitycons}
    \partial_t\left(n^{(\lambda)}\right)+\nabla \left(n^{(\lambda)}\bar{\vct v}^{(\lambda)}\right)=0\,.
\end{align}
Further by multiplying Eq.~\eqref{fp_x1x2} by $m v^{(\lambda)}_i$ and integrating, we can derive the momentum balance equation for the separate momenta of the dumbbell component $\lambda$. In the process we have to calculate the quantity
\begin{align}
    Q^{(\lambda)}  = \int_{\vct r^{(\alpha)}, \vct r^{(\beta)}, \vct v^{(\alpha)}, \vct v^{(\beta)}} F_i(\vct r^{(\beta)} - \vct r^{(\alpha)})  \psi \delta(\vct r- \vct r^{(\lambda)})\,.
\end{align}
After changing the variables to the centre of mass position $\vct x = (\vct r^{(\alpha)}+\vct r^{(\beta)})/2$ and intra-dumbbell vector $\vct R = \vct r^{(\beta)} - \vct r^{(\alpha)}$, we expand the delta function to the linear order in $\vct R$, obtaining
\begin{align}
\begin{split}
    Q^{(\lambda)} & \approx   \int_{\vct x, \vct R, \vct v, \vct v_R} F_i(\vct R) \psi \left[\delta(\vct r- \vct x)- 
(-1)^{\lambda} \vct{R}/2\nabla_r\delta(\vct r- \vct x)\right] = \\
    & = \int_{\vct x, \vct R, \vct v, \vct v_R} F_i(\vct R) \psi \delta(\vct r- \vct x) -  
(-1)^{\lambda}  \frac{1}{2}\nabla_j \int_{\vct x, \vct R, \vct v, \vct v_R} R_j F_i(\vct R) \psi \delta(\vct r- \vct x)\,. 
\end{split}
\end{align}
The expansion above is possible assuming that the distribution function does not vary considerably on the length of a typical dumbbell. Integrating Eq.~\eqref{fp_x1x2} with $m v^{(\lambda)}_i$ leads to
\begin{align}
    \partial_t \left(m n^{(\lambda)} \bar{v}^{(\lambda)}_i\right) + \partial_j\left[m\langle v^{(\lambda)}_i v^{(\lambda)}_j \rangle^{(\lambda)}-\frac{1}{2}\int_{\vct R, \vct v, \vct v_R} R_j F_i(\vct R) \psi\right] =
    - \zeta n^{(\lambda)} (\bar{v}_i^{(\lambda)}-  u^\mathrm{f}_i) - (-1)^{(\lambda)} \int_{ \vct R, \vct v, \vct v_R} F_i(\vct R) \psi\,.
\label{ns_eq_dumbbells}
\end{align}
The gradient term on the left hand side of Eq.~\eqref{ns_eq_dumbbells} can be rewritten as 
\begin{align}
  m\langle v^{(\lambda)}_i v^{(\lambda)}_j \rangle^{(\lambda)}-\frac{1}{2}\int_{\vct R, \vct v, \vct v_R} R_j F_i(\vct R) \psi \equiv  mn^{(\lambda)}\bar{v}^{(\lambda)}_i\bar{v}^{(\lambda)}_j +   T_{ij}^{(\lambda),(\mathrm{k})}+T_{ij}^{(\lambda),(\mathrm{s})} \, , 
\end{align}
where
\begin{subequations}
\begin{align}
    T_{ij}^{(\lambda),(\mathrm{k})} & = m\left\langle \left(\bar{v}^{(\lambda)}_i-v^{(\lambda)}_i\right) \left(\bar{v}^{(\lambda)}_j-v^{(\lambda)}_j\right) \right\rangle^{(\lambda)}, \\
    T_{ij}^{(\lambda),(\mathrm{s})} &= -\frac{1}{2}\int_{\vct R, \vct v, \vct v_R} R_j F_i(\vct R) \psi\,.
\end{align}    
\end{subequations}
$  T_{ij}^{(\lambda),(\mathrm{k})}$ is a part of the stress tensor that describes kinematic transfer of momentum, and can be expanded as a dumbbell pressure term plus viscous corrections which can be ignored when the dumbbells are dilute. With this assumption we have
\begin{align}
     T_{ij}^{(\lambda),(\mathrm{k})}  \approx  p^{(\lambda )}  \delta_{ij}  \,, ~~   p^{(\lambda )}   =  n^{(\lambda)} k T \, . 
\end{align}
$T_{ij}^{(\lambda),(\mathrm{s})} $ describes the stress induced by contractions of the springs and its corresponding evolution equation will be derived in App.~\ref{app:modelderivation}. Using Eq.~\eqref{eq:densitycons}, Eq.~\eqref{ns_eq_dumbbells} then turns into
\begin{align}
     m n^{(\lambda)} \frac{\mathcal D}{\mathcal D t }  \bar{v}^{(\lambda)}_i  + \partial_j T_{ij}^{(\lambda),(\mathrm{s})}   + \partial_i p^{(\lambda )} = - (-1)^{(\lambda)} \int_{ \vct R, \vct v, \vct v_R} F_i(\vct R) \psi 
    - \zeta n^{(\lambda)}(\bar{v}_i^{(\lambda)}-u^\mathrm{f}_i) \,, 
\label{ns_eq_dumbbells1}
\end{align}
where 
\begin{align}
    \frac{\mathcal D}{\mathcal D t }  \bar{v}^{(\lambda)}_i     = \partial_t  \bar{v}^{(\lambda)}_i + \bar{v}^{(\lambda)}_j \partial_j    \bar{v}^{(\lambda)}_i\, . 
\end{align}
Because we take the dumbbells to be dilute and their motion to be overdamped by the background fluid, we can set the first term on the left hand side of Eq. (\ref{ns_eq_dumbbells1}) to zero. Then we have
\begin{align}
   \partial_i p^{(\lambda)}+\partial_j  T_{ij}^{(\lambda),(\mathrm{s})} =
    - \zeta n^{(\lambda)} (\bar{v}_i^{(\lambda)}-  u^\mathrm{f}_i) - (-1)^{(\lambda)} \int_{ \vct R, \vct v, \vct v_R} F_i(\vct R) \psi\,.
\label{ns_eq_dumbbells12}
\end{align}
There are two terms in Eq.~\eqref{ns_eq_dumbbells12} which prevent the momentum from being conserved. Firstly, there is the term coming from the averaged dumbbell force $F_i(\vct R)$ evaluated at the dumbbell center of mass $\vct r$. Because the beads act on each other with equal and opposite forces, the contribution of this term will vanish when the equations for the two dumbbell components are added. Next, the term in Eq.~\eqref{ns_eq_dumbbells12} that represents drag of the background fluid expresses coupling with the solvent, together with which the dumbbells form a \quotes{two-fluid medium} \cite{twolfuid1,twolfuid2,twolfuid3}. To see that the total linear momentum is conserved, we consider the conservation laws of the background fluid. The equations of motion of the background fluid are
\begin{subequations}  \label{eq:fluidequations}
\begin{align} \label{eq:densityfluid}
    \partial_t\rho^\mathrm{f} + \partial_i (\rho^\mathrm{f}  u_i^\mathrm{f}) & = 0\,,\\  \label{eq:momentumequation}
   \rho^\mathrm{f} \frac{\mathcal D}{\mathcal D t } u_i^\mathrm{f} +  \partial_j T^{(\mathrm{f})}_{ij} &= - 
 \zeta\left( 2  n^\mathrm{d} u_i^\mathrm{f}-  \sum_{\lambda = \alpha,\beta} n^\mathrm{(\lambda)} 
 \bar{v}_i^{(\lambda)}\right)\,,   
 \end{align}    
\end{subequations}
where
\begin{align}  \label{eq:fluidstress}
    T^{(\mathrm{f})}_{ij} = p^\mathrm{f}  \,   \delta_{ij} - \eta^\mathrm{f}_{ij k l } D^\mathrm{f}_{ k l  } \, . 
\end{align}
$\eta^\mathrm{f}_{ij k l }$ is the viscosity tensor of the background fluid. After adding Eq.~\eqref{ns_eq_dumbbells12} for $\alpha$ and $\beta$ to Eq.~\eqref{eq:momentumequation}, the conservation of the total momentum becomes exact, i.e., we have
\begin{equation} 
   \rho^\mathrm{f} \frac{\mathcal D}{\mathcal D t } u^{\mathrm{f} }_i + \partial_i p^{\mathrm{d}}+\partial_j   T_{ij}^{(\mathrm{s})} +  \partial_j T^{(\mathrm{f})}_{ij} = 0 \,       
\end{equation}
with
\begin{subequations}
\begin{align}
    p^{\mathrm{d}}&=2n^{\mathrm{d}}kT\,,\\
    T_{ij}^{(\mathrm{s})} &= - \int_{\vct R, \vct v, \vct v_R} R_j F_i(\vct R) \psi\,.
\end{align}    
\end{subequations}

\section{Derivation of the rheological equations}
\label{app:modelderivation}

In this Appendix we derive rheological equations for a broad class of dumbbell models, in which in addition to the dumbbells displaying even and odd elastic forces, they can be subject to an internally or externally induced rotation. The force acting on the beads is assumed to take the form
\begin{equation}
     F_i (\vct R)  = \delta R \left(\kappa_e \delta_{ij} + \frac{R_*}{R} \kappa_o\varepsilon_{ij} \right)\hat{r}_j+\frac{\zeta\Omega}{2}~\frac{R_*^2}{R}\varepsilon_{ij}\hat{r}_j+\alpha\frac{\zeta}{2}\frac{R_*^2}{R}\left(\hat{r}_j \partial_j u_i - \hat{r}_i\hat{r}_l\hat{r}_j\partial_ju_l\right)\,.
     \label{force_general_long}
\end{equation}
The terms in Eq. (\ref{force_general_long}) have the following interpretation:
\begin{itemize}
    \item $\delta R \left(\kappa_e \delta_{ij} + \frac{R_*}{R} \kappa_o\varepsilon_{ij} \right)\hat{r}_j$ represents both the even and odd elastic forces acting on the dumbbell that are proportional to the extension of the spring.
    \item $\frac{\zeta\Omega}{2}~\frac{R_*^2}{R}\varepsilon_{ij}\hat{r}_j$ is a force that induces an approximately steady rotation of the dumbbell with frequency $\Omega$ in the overdamped limit. It is related to a nonzero torque exerted either by internal degrees of freedom, or by an external driving force.
    \item $\alpha\frac{\zeta}{2}\frac{R_*^2}{R}\left(\hat{r}_j \partial_j u_i - \hat{r}_i\hat{r}_l\hat{r}_j\partial_ju_l\right)$ is a nonconventional term, which, when $\alpha$ is set to 1, cancels the component of the fluid drag perpendicular to $\vct R$ in the equations of motion. Namely, setting $\alpha=1$ and $\kappa_o=0$ leads to $\dot{R}_i \approx -\Omega \varepsilon_{ij} R_j + \hat{r}_i\hat{r}_lR_j\partial_j u_l -\delta R \kappa_e \hat{r}_i + \frac{1}{\zeta} f_i^1$, which ensures that the angular motion of the dumbbells remains locked to the rotating magnetic field in agreement with the findings of Ref.~\cite{Soni2019}, modulo fluctuations. On the contrary, for models without external driving there is no mechanism that could lead to such a locking, and consequently $\alpha=0$.
\end{itemize}
We now proceed to derive the rheological equations. The main object of interest is the stress tensor $T_{ij}^{(\mathrm{s})}$ defined in Eq.~\eqref{eq:Ts}, which exhibits a nontrivial time evolution that can be found using the Fokker-Planck equation, Eq.~\eqref{fp_eq}. We simplify the resulting equations by performing simultaneously two expansions: the first one is expanding the equations up to the first order in gradients around an isotropic steady-state solution in order to obtain a hydrodynamic description; and the second is expanding terms contributing to the stress tensor up to the second order in $\delta R/R_*$, and then replacing every instance of $\delta R^2$ by its average value given in Eq.~\eqref{eq:R2_average}. The validity of these approximations is ultimately justified by the agreement we obtain with the numerical simulations.

Let us begin by decomposing the stress tensor $T_{ij}^{(\mathrm{s})}$ given in Eq.~\eqref{eq:Ts} into separate terms. To this end, we define three symmetric tensors $C_{ij}$, $X_{ij}$ and $Y_{ijkl}$:
\begin{equation}
    C_{ij} \equiv R_* \int_{\vct R}\hat{r}_i\hat{r}_j\delta R\psi\,, ~~~~X_{ij} \equiv R_*^2 \int_{\vct R} \hat{r}_i\hat{r}_j\psi\,, ~~~~Y_{ijkl} \equiv R_*^2\int_{\vct R} \hat{r}_i\hat{r}_j\hat{r}_k\hat{r}_l\psi \,,
    \label{CXY_define}
\end{equation}
and notice that 
\begin{equation}
    \int_{\vct R}\hat{r}_iR_j\delta R\psi = R_*\int_{\vct R}\hat{r}_i\hat{r}_j\delta R\psi + \int_{\vct R}\hat{r}_i\hat{r}_j\delta R^2\psi \approx C_{ij} + \frac{kT}{\kappa_e R_*^2}X_{ij} \,.
\end{equation}
$T_{ij}^{(\mathrm{s})}$ can then be decomposed as
\begin{align}
    T_{ij}^{(\mathrm{s})} = -\kappa_{ij k l } C_{k l } - \Omega_{ijkl} X_{kl} - \alpha\frac{\zeta}{2}\partial_ku_iX_{kj} + \alpha\frac{\zeta }{2}\partial_ku_lY_{ijkl}
    \label{eq:Tsdefinition21}
\end{align}
with $\kappa_{ijkl}$ defined as
\begin{equation}
  \kappa_{ij k l }  =     \kappa_e \gamma_{i j kl } +  \frac{1}{2}  \kappa_e \delta_{ij} \delta_{k l }  + \kappa_{o} \gamma^{o}_{i j k l }  +\frac{1}{2}  \kappa_{o} \varepsilon_{ij } \delta_{k l }
  \label{eq:kappaijkl}
\end{equation}
and $\Omega_{ijkl}$ as
\begin{equation}
    \Omega_{ij k l }  = \frac{kT}{R_*^2} \gamma_{i j kl } +  \frac{1}{2}  \frac{kT}{R_*^2} \delta_{ij} \delta_{k l }+\frac{\zeta}{2} \Omega~\gamma^{o}_{i j k l }  +\frac{\zeta}{4} \Omega~\varepsilon_{ij } \delta_{k l }\,. 
    \label{eq:omegaijkl}
\end{equation}
Firstly we will obtain steady-state expressions for the tensors $C_{ij}$, $X_{ij}$ and $Y_{ijkl}$. Due to isotropy of the system in equilibrium and symmetricity of the tensors they have to take the form $C_{ij} = C \delta_{ij} + \mathcal{O}(\partial)$, $X_{ij} = X \delta_{ij} + \mathcal{O}(\partial)$, $Y_{ijkl} = Y\left(\delta_{ij}\delta_{kl}+\delta_{ik}\delta_{jl}+\delta_{il}\delta_{jk}\right)+ \mathcal{O}(\partial)$, where $C$, $X$, $Y$, are some coefficients. These coefficients can be determined as follows:
\begin{equation}
 \frac12 X_{ii} = \frac{R_*^2}{2} \int_{\vct R}\psi = \frac{n^\mathrm{d}R_*^2}{2} = X \,,
 \label{X_average}
\end{equation}
\begin{equation}
  \frac18 Y_{iikk} = \frac{R_*^2}{8} \int_{\vct R}\psi = \frac{n^\mathrm{d}R_*^2}{8} =  Y \,.
\end{equation}
In fact the calculation above shows that the trace of $X_{ij}$ is fixed by the definition of $X_{ij}$, and it is only the traceless component of $X_{ij}$ that can evolve out of equilibrium. To find the steady-state expression for $C_{ij}$, Eq.~\eqref{fp_eq} is multiplied by $\delta R$ and integrated, producing in the absence of velocity gradients ($\partial_i u_j=0$):
\begin{equation}
   \frac{\mathcal{D}}{\mathcal{D} t } \int_{\vct R}\delta R~\psi 
   =   -\frac{2 \kappa_e}{\zeta}  \int_{\vct R}\delta R~\psi+\frac{2kT}{\zeta}\int_{\vct R}\frac{1}{R}\psi \approx -\frac{2 \kappa_e}{\zeta}  \int_{\vct R}\delta R~\psi+\frac{2n^\mathrm{d}kT}{\zeta R_*}
    \label{eq:dR_evol}
\end{equation}
with the approximation valid up to the first order in $kT/(\kappa_e R_*^2)$. It implies that
\begin{equation}
    \frac12 C_{ii} = \frac{R_*}{2} \int_{\vct R}\delta R~\psi = \frac{n^\mathrm{d}kT}{2 \kappa_e} = C \,.
    \label{C_average}
\end{equation}
Altogether, $T_{ij}^{(\mathrm{s})}$ can be expanded up to the first order in gradients as follows,
\begin{subequations}
\begin{align} \label{eq:Ts_define_0}
    T^{(\mathrm{s})}_{ij} & = -n^\mathrm{d} k T\delta_{ij} - \frac{\kappa_o}{2\kappa_e}n^\mathrm{d} k T\varepsilon_{ij}- \alpha \frac{n^\mathrm{d} R_*^2 \zeta}{8}D_{\langle ij \rangle}-\frac{n^\mathrm{d} R_*^2 \zeta}{8}\left(2\Omega-\alpha D^A\right)\varepsilon_{ij} + \tilde{T}^{(\mathrm{s})}_{ij} + \mathcal{O}(\partial^2)\,,\\
    \tilde{T}^{(\mathrm{s})}_{ij} & = -\kappa_{ijkl}\tilde{C}_{kl} - \Omega_{ijkl}\tilde{X}_{kl}\,, \label{eq:Ts_define}
\end{align} 
\end{subequations}
with $\tilde{C}_{ij}, \tilde{X}_{ij} \sim \mathcal{O}(\partial)$. Following Appendix \ref{app:stressderivation}, the total stress tensor is obtained by adding the fluid part of the stress tensor and the dumbbell pressure term, resulting in 
\begin{equation}
    T_{ij} = \left(p^\mathrm{f}+n^\mathrm{d} kT\right)\delta_{ij} - \frac{\kappa_o}{2 \kappa_e}n^\mathrm{d} k T\varepsilon_{ij}-\eta_{ijkl}D_{kl}- \alpha \frac{n^\mathrm{d} R_*^2 \zeta}{8}D_{\langle ij \rangle}-\frac{n^\mathrm{d} R_*^2 \zeta}{8}\left(2\Omega-\alpha D^A\right)\varepsilon_{ij} + \tilde{T}^{(\mathrm{s})}_{ij} + \mathcal{O}(\partial^2)\,.
    \label{eq:Ts_long}
\end{equation}
Evidently, on this level of approximation, the effect of the \quotes{drag-cancelling} term parametrized by $\alpha$ in Eq. (\ref{force_general_long}) is to increase the shear viscosity of the fluid-dumbbell system and to introduce a rotational viscosity equal to $\eta_R = n^\mathrm{d} \zeta R_*^2/8$. Note also that the constant anti-symmetric stress disappears when $\zeta\Omega = -2 kT\kappa_o/(\kappa_eR_*^2)$.

In order to find the dynamical equations for $\tilde{T}_{ij}^{(\mathrm{s})}$, Eq.~\eqref{fp_eq} is multiplied by $\hat{r}_i \hat{r}_j\delta R$ and after integrating one obtains the equation
\begin{equation}
    \begin{split}
        \frac{\mathcal{D}}{\mathcal{D} t}C_{ij} =& (1-\alpha)\partial_l u_i C_{lj}+ (1-\alpha)\partial_l u_j C_{li}+\partial_l u_m R_*\int_{\vct R} \hat{r}_i\hat{r}_j\hat{r}_l\hat{r}_m \left(R_*-\delta R+2\alpha\delta R - \alpha \frac{4kT}{\kappa_e R_*}\right)\psi -\frac{2\kappa_e}{\zeta}C_{ij} \\
        & -\frac{4kT}{\kappa_e R_*^2}\left(\frac{\kappa_o}{\zeta}+\Omega\right)X_{\langle ij\rangle}^o  -2\Omega C_{\langle ij \rangle}^o+\alpha\frac{2kT}{\kappa_e R_*^2}\left(\partial_l u_i X_{lj}+\partial_l u_j X_{li}\right)-\frac{8kT}{\zeta R_*^2}C_{\langle ij \rangle}+\frac{2kT}{\zeta R_*^2}X_{ij} \,,
    \end{split}
    \label{eq:Cij0}
\end{equation}
where we expanded in $\delta R/R_*$ up to second order and replaced $\delta R^2$ by its equilibrium value $kT/\kappa_e$. In the derivation we used the identities
\begin{subequations}
\begin{align}
    \partial_{R_l} (\hat{r}_i \hat{r}_j\delta R) & = \left(\delta_{il}\hat{r}_j+\delta_{jl}\hat{r}_i-2 \hat{r}_i\hat{r}_j\hat{r}_l\right)\frac{\delta R}{R_*} + \hat{r}_i\hat{r}_j\hat{r}_l\,, \\
    \label{eq:quadratic}
    \partial_{R_l}^2 (\hat{r}_i \hat{r}_j\delta R) & = 2\left(\delta_{ij}-2\hat{r}_i\hat{r}_j\right)\frac{\delta R}{R}+\frac{1}{R}\hat{r}_i\hat{r}_j\,.
\end{align}    
\end{subequations}
The third term on the right-hand side of Eq. (\ref{eq:Cij0}) can be simplified as follows,
\begin{equation}
    \partial_l u_m R_*\int_{\vct R} \hat{r}_i\hat{r}_j\hat{r}_l\hat{r}_m \left(R_*-\delta R+2\alpha\delta R - \alpha \frac{4kT}{\kappa_e R_*}\right)\psi = \partial_l u_m \frac{n^\mathrm{d}R_*^2}{8}\left(\delta_{ij}\delta_{lm}+\delta_{il}\delta_{jm}+\delta_{im}\delta_{jl}\right)+\mathcal{O}(\partial^2)+\mathcal{O}(kT \cdot \partial)\,.
\end{equation}
Dropping all terms of the order $\mathcal{O}(\partial^2), \mathcal{O}(kT\cdot\partial), \mathcal{O}((kT)^2)$, Eq. (\ref{eq:Cij0}) can be transformed into
\begin{equation}
    \frac{\partial}{\partial t} C_{ij} = \frac{n^\mathrm{d}R_*^2 }{4}\left(\delta_{ij}D_{ll} + D_{\langle ij \rangle}\right)-\frac{2\kappa_e}{\zeta}C_{ij}-2 \Omega C_{\langle ij \rangle}^o+\frac{2 k T}{\zeta R_*^2}X_{ij}-\frac{4kT}{\kappa_e R_*^2}\left(\frac{\kappa_o}{\zeta}+\Omega\right)X_{\langle ij \rangle}^o\,.
    \label{Cij_evolution}
\end{equation}
Similarly, multiplying Eq.~\eqref{fp_eq} by $\hat{r}_i \hat{r}_j$ and integrating, one obtains the equation for the evolution of $X_{ij}$. Dropping terms of the order $\mathcal{O}(\partial^2), \mathcal{O}(kT\cdot\partial), \mathcal{O}((kT)^2)$, one obtains
\begin{equation}
\begin{split}
      \frac{\partial}{\partial t}X_{ij} = &(1-\alpha)\left\{\partial_mu_iX_{mj} + \partial_m u_j X_{mi} - 2\partial_mu_lR_*^2\int_{\vct R} \hat{r}_i\hat{r}_j\hat{r}_l\hat{r}_m \psi\right\} \\
      &-4\left(\frac{\kappa_o}{\zeta}-\Omega\right)C_{\langle ij \rangle}^o-\frac{8kT}{\zeta R_*^2}X_{\langle ij \rangle}-2\left(\Omega-\frac{4kT\kappa_o}{\zeta \kappa_e R_*^2}\right)X_{\langle ij \rangle}^o\,,
      \label{eq:xevol_long0}
\end{split}
\end{equation}
which can be further simplified as
\begin{equation}
      \frac{\partial}{\partial t}X_{ij} = (1-\alpha)\frac{n^\mathrm{d}R_*^2}{2}D_{\langle ij \rangle} 
      -4\left(\frac{\kappa_o}{\zeta}-\Omega\right)C_{\langle ij \rangle}^o-\frac{8kT}{\zeta R_*^2}X_{\langle ij \rangle}-2\left(\Omega-\frac{4kT\kappa_o}{\zeta \kappa_e R_*^2}\right)X_{\langle ij \rangle}^o\,.
      \label{eq:xevol_long}
\end{equation}
Using the definition of $\tilde{T}^{(\mathrm{s})}_{ij}$ given in Eq. (\ref{eq:Ts_define}) and Eqs. (\ref{Cij_evolution}) and (\ref{eq:xevol_long}), the equation of motion for the trace component turns out to be
\begin{equation} \label{eq:long_sym}
     \frac{\partial}{\partial t}\tilde{T}^{(\mathrm{s})} + \frac{2 \kappa_e}{\zeta}\tilde{T}^{(\mathrm{s})} = - \frac{n^\mathrm{d} R_*^2 \kappa_e}{2} D \,, 
\end{equation}
and the antisymmetric component is simply proportional to the trace part,
\begin{equation} \label{eq:long_antisym}
    \tilde{T}^{(\mathrm{s})A} =  \frac{\kappa_o}{\kappa_e}\tilde{T}^{(\mathrm{s})}\,.
\end{equation}
Regarding the traceless component $\tilde{T}^{(\mathrm{s})}_{\langle ij \rangle}$, the set of two coupled first-order equations of motion given in Eqs. (\ref{Cij_evolution}) and (\ref{eq:xevol_long}) can be expressed as a second-order equation of motion for $\tilde{T}^{(\mathrm{s})}_{\langle ij \rangle}$. In order to achieve this, Eq. (\ref{Cij_evolution}) multiplied by $-\kappa_{ijkl}$ and Eq. (\ref{eq:xevol_long}) multiplied by $-\Omega_{ijkl}$ are added together, and in the resulting equations $\tilde{C}_{\langle ij \rangle}$ is replaced by $\tilde{C}_{\langle ij \rangle} = -\frac{1}{\kappa_e^2+\kappa_o^2}\left(\kappa_e \gamma_{ijkl} - \kappa_o \gamma_{ijkl}^o\right)\left(\tilde{T}^{(\mathrm{s})}_{\langle kl \rangle}+\Omega_{klmn}\tilde{X}_{mn}\right)$. This produces a system of equations of the form
\begin{subequations}
\begin{align}
    \frac{\partial}{\partial t}\tilde{T}^{(\mathrm{s})}_{\langle ij \rangle} & = \alpha_{ijkl}\tilde{T}^{(\mathrm{s})}_{\langle kl \rangle} + \beta_{ijkl}\tilde{X}_{\langle kl \rangle} + \mu_{ijkl}D_{\langle kl\rangle}\,, \\
    \frac{\partial}{\partial t}\tilde{X}_{\langle ij \rangle} & = \zeta_{ijkl}\tilde{T}^{(\mathrm{s})}_{\langle kl \rangle} + \eta_{ijkl}\tilde{X}_{\langle kl \rangle} + \nu_{ijkl}D_{\langle kl\rangle}\,,
\end{align}
\end{subequations}
where $\alpha_{ijkl}$, $\beta_{ijkl}$, $\zeta_{ijkl}$, $\eta_{ijkl}$, $\mu_{ijkl}$, $\nu_{ijkl}$ are certain tensors which we do not write explicitly due to the lengthiness of the formulas, which are at the same time not particularly illuminating. This system of equations is then reformulated as
\begin{equation}
    \frac{\partial^2}{\partial t^2}\tilde{T}^{(\mathrm{s})}_{\langle ij \rangle} - \left(\alpha_{ijkl}+\eta_{ijkl}\right)\frac{\partial}{\partial t}\tilde{T}^{(\mathrm{s})}_{\langle kl \rangle} - \left(\beta_{ijkl}\zeta_{klmn}-\eta_{ijkl}\alpha_{klmn}\right)\tilde{T}^{(\mathrm{s})}_{\langle mn \rangle} = \mu_{ijkl}\frac{\partial}{\partial t}D_{\langle kl\rangle} + \left(\beta_{ijkl}\nu_{klmn}-\eta_{ijkl}\mu_{klmn}\right)D_{\langle mn\rangle}\,.
    \label{eq:Tevol_long}
\end{equation}
This equation has the same form as Eq. (\ref{eq:general_jeffreys}).

In the end, the evolution of the stress tensor can be described by a set of second-order differential equations. However, in many cases of interest one is able to make approximations which reduce the problem to a set of much simpler first-order equations. In what follows we shall focus on the odd-dumbbell model introduced in the main text and we assume that the elastic relaxation rate $\kappa_e/\zeta$ and the driving frequency $\Omega$ obey the inequality $\kappa_e/\zeta \gg \Omega$. This introduces a separation of timescales, and the behavior of the system is qualitatively different at short times $t\lessapprox \zeta/\kappa_e$ and at longer times $t\gg \zeta/\kappa_e$. Lets us take a closer look at these two regimes.

\subsection{Short-time regime}

This regime describes the behavior at times $t \lessapprox \zeta/\kappa_e$, in which case we approximate $\frac{\partial}{\partial t} \approx \kappa_e/\zeta$. The various terms in Eqs. (\ref{Cij_evolution}) and (\ref{eq:xevol_long}) can be grouped according to their magnitude, noticing that $\tilde{C}_{ij}\sim\tilde{X}_{ij}\sim\mathcal{O}(\partial)$ and $\kappa_e/\zeta\gg\Omega$, $\kappa_e/\zeta \gg kT/(\zeta R_*^2)$. We observe that the majority of the terms can be neglected due to their smallness, and the resultant equations can be rewritten using the same procedure as the one used to obtain Eq.~\eqref{eq:Tevol_long}. The result is
\begin{subequations}
\begin{align}
   \frac{\partial}{\partial t}\left\{\frac{\partial}{\partial t} \tilde{T}^{(\mathrm{s})}_{\langle  i j  \rangle } + \frac{2\kappa_e}{\zeta}  \tilde{T}^{(\mathrm{s})}_{\langle ij  \rangle }+\frac{n^\mathrm{d}R_*^2  \kappa_e}{4} D_{\langle ij \rangle }
     +   \frac{n^\mathrm{d}R_*^2 \kappa_o}{4}D^o_{\langle ij \rangle }\right\}& = 0\,, \\
     \frac{\partial}{\partial t}\tilde{T}^{(\mathrm{s})} + \frac{2 \kappa_e}{\zeta}\tilde{T}^{(\mathrm{s})} + \frac{n^\mathrm{d}R_*^2 \kappa_e}{2} D &=0\,,\\
      \tilde{T}^{(\mathrm{s})A} -  \frac{\kappa_o}{\kappa_e}\tilde{T}^{(\mathrm{s})} &=0\,.
\end{align} \label{eq:Tevol_elastic_0}
\end{subequations}
The first of these equations can be rewritten with the help of an integration constant $\frac{2\kappa_e}{\zeta}\tilde{\tau}_{\langle ij \rangle}$, which physically corresponds to the magnitude of $\tilde{T}^{(\mathrm{s})}_{\langle ij \rangle}$ after elastic relaxation:
\begin{equation}
   \frac{\partial}{\partial t} \tilde{T}^{(\mathrm{s})}_{\langle  i j  \rangle } + \frac{2\kappa_e}{\zeta}  \left(\tilde{T}^{(\mathrm{s})}_{\langle  i j  \rangle }-\tilde{\tau}_{\langle ij\rangle}\right) = -\frac{R_*^2 n^\mathrm{d} \kappa_e}{4} D_{\langle ij \rangle }
     -   \frac{R_*^2 n^\mathrm{d}\kappa_o}{4}D^o_{\langle ij \rangle }\,. \label{eq:Tevol_elastic_01}
\end{equation}
By including also the viscous part of the stress tensor as seen in Eq. (\ref{eq:Ts_long}) and setting $\alpha=0$ we recover the odd Jeffreys model with coefficients given in Eq. (\ref{eq:coeffs_1}). The constant stress $\tilde{\tau}_{ij}$ cannot be determined from Eq. (\ref{eq:Tevol_elastic_01}) alone; still, it is possible to find it by with the help of Eq. (\ref{Cij_evolution}), from which follows that at times $t \gg \zeta/\kappa_e$ $\tilde{C}_{\langle ij\rangle}$ relaxes to 
\begin{equation}
    \tilde{C}_{\langle ij \rangle} \approx \frac{n^\mathrm{d}R_*^2\zeta}{8 \kappa_e}D_{\langle ij \rangle}+\frac{kT}{\kappa_e R_*^2}\tilde{X}_{\langle ij \rangle}-\frac{2kT\kappa_o}{\kappa_e^2 R_*^2}\tilde{X}_{\langle ij \rangle}^o.
    \label{eq:c_relaxed}
\end{equation}
From the definition of $\tilde{T}_{ij}^{(\mathrm{s})}$, Eq. (\ref{eq:Ts_define}), follows that
\begin{equation}
    \tilde{\tau}_{\langle ij \rangle} = -\frac{n^\mathrm{d}R_*^2\zeta}{8}D_{\langle ij \rangle}-\frac{\kappa_o}{\kappa_e}\frac{n^\mathrm{d}R_*^2\zeta}{8} D_{\langle ij \rangle}^o-\frac{2kT}{R_*^2}\left(1+\frac{\kappa_o^2}{\kappa_e^2}\right)\tilde{X}_{\langle ij \rangle}-\left(\frac{\zeta\Omega}{2}-\frac{kT\kappa_o}{R_*^2\kappa_e}\right)\tilde{X}_{\langle ij \rangle}^o\,.
    \label{eq:tau_full}
\end{equation}
For the purposes of analysing the numerical results, we will be interested in the value of the correlator $\langle \tilde{\tau}_{\langle xx \rangle}\tilde{\tau}_{\langle xx \rangle} \rangle$ in the situation when $D_{ij}=0$. Using the fact that the vector $R_i$ is isotropically distributed in the steady state, one obtains
\begin{equation}
    \langle \tilde{\tau}_{\langle xx \rangle}\tilde{\tau}_{\langle xx \rangle} \rangle =  \frac{n^\mathrm{d} R_*^4 \zeta\Omega}{32}\left(\zeta\Omega-\frac{4kT\kappa_o}{ R_*^2\kappa_e}\right) +\mathcal{O}\left((kT)^2\right)\,.
    \label{eq:stress_corr_relaxed}
\end{equation}
It is also clear from Eqs.~\eqref{eq:Tevol_elastic_0} that after $t\gg \zeta/\kappa_e$ both the trace and the anti-symmetric component of the stress tensor relax to their equilibrium value of 0, provided $D=0$.

\subsection{Long-time regime}

This regime describes the behavior at times $t \gg \zeta/\kappa_e$. The trace and the antisymmetric components of $\tilde{T}^{(\mathrm{s})}_{ij}$ relax to
\begin{equation}
    \tilde{T}^{(\mathrm{s})} = -\frac{n^\mathrm{d}R_*^2\zeta}{4}D\,, \quad\quad \tilde{T}^{(\mathrm{s})A} = -\frac{\kappa_o}{\kappa_e}\frac{n^\mathrm{d}R_*^2\zeta}{4}D\,,   
\end{equation}
which contributes to the total fluid viscosity. This relaxation is related to the fact that only the traceless component of $X_{ij}$ can have a non-zero value out of equilibrium. In the long-time regime $\tilde{C}_{\langle ij \rangle}$ is given by the formula Eq.~\eqref{eq:c_relaxed}, which together with the definition of $\tilde{T}^{(\mathrm{s})}_{ij}$ in Eq.~(\ref{eq:Ts_define}) and the formula for the evolution of $X_{ij}$ in Eq.~(\ref{eq:xevol_long}) produces the odd Jeffreys model
\begin{equation}
    \frac{\partial}{\partial t} \tilde{T}^{(\mathrm{s})}_{\langle  i j  \rangle } 
      +   \chi_s  \tilde{T}^{(\mathrm{s})}_{\langle ij  \rangle }   + \chi_{o}  \tilde{T}^{(\mathrm{s})o}_{\langle ij  \rangle }
    =  - \gamma_s \frac{\partial}{\partial t} D_{\langle  i j  \rangle } - \gamma_o \frac{\partial}{\partial t} D^o_{\langle  i j  \rangle } -   \left(\zeta_s +\gamma_s\chi_s-\gamma_o\chi_o \right)  D_{\langle ij \rangle } -   \left(\zeta_s +\gamma_o\chi_s+\gamma_s\chi_o \right)  D^o_{\langle ij \rangle }
    \label{eq:long_time_full}
\end{equation}
with coefficients
\begin{equation}
\begin{split}
    \chi_s & = \frac{8kT}{\zeta R_*^2}\left(1+\frac{\kappa_o^2}{\kappa_e^2}\right)\,, \quad\quad\quad \gamma_s = \frac{n^\mathrm{d}R_*^2\zeta}{8}\,, \quad\quad\quad \zeta_s = (1-\alpha) \gamma_s\chi_s+\gamma_o\chi_o\,, \\
    \chi_o & = 2\Omega-\frac{4kT \kappa_o}{\kappa_e R_*^2\zeta}\,,   \quad\quad\quad\quad \gamma_o = \frac{\kappa_o}{\kappa_e}\frac{n^\mathrm{d}R_*^2\zeta}{8}\,, \quad\quad \zeta_o = (1-\alpha) \gamma_s\chi_o-\gamma_o\chi_s\,.
\end{split}
\end{equation}
By including also the viscous part of the stress tensor as seen in Eq.~(\ref{eq:Ts_long}) and setting $\alpha=0$ we recover the odd Jeffreys model described with coefficients given in Eq.~(\ref{eq:coeffs_2}).

Let us conclude this section by determining the asymptotic late-time form of the stress-energy tensor for a constant flow $D_{ij} = \mathrm{const.}$ This amounts to setting $\frac{\partial}{\partial t} \tilde{T}^{(\mathrm{s})}_{\langle  i j  \rangle }=\frac{\partial}{\partial t} D_{\langle  i j  \rangle }=0$ in Eq.~\eqref{eq:long_time_full}, solving for $\tilde{T}_{ij}^{(\mathrm{s})}$, and substituting the result in Eq.~\eqref{eq:Ts_define_0}. The result is
\begin{equation} \label{eq:late_time}
    T^{(\mathrm{s})}_{ij} = -n^\mathrm{d} k T\delta_{ij} - \frac{\kappa_o}{2\kappa_e}n^\mathrm{d} k T\varepsilon_{ij}- \frac{n^\mathrm{d} R_*^2 \zeta}{4}D_{\langle ij \rangle}-\frac{n^\mathrm{d} R_*^2 \zeta}{8}\left(2\Omega-\alpha D^A\right)\varepsilon_{ij} + \mathcal{O}(\partial^2)\,.
\end{equation}

\end{document}